\documentclass[lettersize,journal]{IEEEtran}

\usepackage{algorithm}
\usepackage{algpseudocode}
\usepackage{amsmath,amssymb,amsfonts}
\usepackage{booktabs}
\usepackage{cite}
\usepackage[acronym]{glossaries}
\usepackage{graphicx}
\usepackage{hyperref}
\usepackage{listings}
\usepackage{pifont}
\usepackage{subfig}
\usepackage{subcaption}
\usepackage{stfloats}
\usepackage{tabulary}
\usepackage{tabularx}
\usepackage{textcomp}
\usepackage{url}
\usepackage{xparse}
\usepackage[table,svgnames]{xcolor} 

\usepackage{glossaries}
\newacronym{pdp}{PDP}{Programmable Data Plane}
\newacronym{pdps}{PDPs}{Programmable Data Planes}
\newacronym{p3po}{P3PO}{Parallel Processing for Priority Ordering}
\newacronym{ppqa}{PPQA}{Parallel PIFO Queuing Architecture}
\newacronym{pifo}{PIFO}{Push-in First-out}
\newacronym{fifo}{FIFO}{First-in First-out}
\newacronym{sppifo}{SPPIFO}{SPPIFO}
\newacronym{aifo}{AIFO}{AIFO}
\newacronym{rifo}{RIFO}{RIFO}
\newacronym{fcts}{FCTs}{Flow Completion Times}
\newacronym{fcts1}{FCTs}{flow completion times}
\newacronym{aqm}{AQM}{Active Queue Management}
\newacronym{stfq}{STFQ}{Start Time Fair Queuing}
\newacronym{qos}{QoS}{Quality of Service}

\definecolor{codegreen}{rgb}{0,0.6,0}
\definecolor{codegray}{rgb}{0.5,0.5,0.5}
\definecolor{codepurple}{rgb}{0.58,0,0.82}
\definecolor{backcolour}{rgb}{0.95,0.95,0.92}

\newif\ifshowblue
\showbluefalse

\ifshowblue
    \newcommand{\blueblock}[1]{{\color{blue}#1}}
\else
    \newcommand{\blueblock}[1]{#1}
\fi

\newif\ifshowgreen
\showgreenfalse

\ifshowgreen
    \newcommand{\greenblock}[1]{{\color{DarkGreen}#1}}
\else
    \newcommand{\greenblock}[1]{#1}
\fi

\newif\ifshowpurple
\showpurplefalse

\ifshowpurple
    \newcommand{\purpleblock}[1]{{\color{codepurple}#1}}
\else
    \newcommand{\purpleblock}[1]{#1}
\fi



\lstdefinestyle{mystyle}{
    backgroundcolor=\color{backcolour},   
    commentstyle=\color{codegreen},
    keywordstyle=\color{magenta},
    numberstyle=\tiny\color{codegray},
    stringstyle=\color{codepurple},
    basicstyle=\ttfamily\footnotesize,
    breakatwhitespace=false,         
    breaklines=true,                 
    captionpos=b,                    
    keepspaces=true,                 
    numbers=left,                    
    numbersep=5pt,                  
    showspaces=false,                
    showstringspaces=false,
    showtabs=false,                  
    tabsize=2
}
\usepackage{array}
\newcolumntype{L}{>{\raggedright\arraybackslash}p{1.1cm}}
\newcolumntype{C}{>{\centering\arraybackslash}p{1.4cm}}

\def\BibTeX{{\rm B\kern-.05em{\sc i\kern-.025em b}\kern-.08em
    T\kern-.1667em\lower.7ex\hbox{E}\kern-.125emX}}
\AtBeginDocument{\definecolor{ojcolor}{cmyk}{0.93,0.59,0.15,0.02}}

\begin{document}
\title{Parallel Architectures For Priority Scheduling In Programmable
  Data Plane Switches} 

\author{Nikhil Shinde\IEEEauthorrefmark{1}, 
    Krishna M. Sivalingam\IEEEauthorrefmark{1} \IEEEmembership{(Fellow, IEEE)} and
    Gauravdeep Shami\IEEEauthorrefmark{2}
\\
\IEEEauthorrefmark{1} {Department of Computer Science and
  Engineering, Indian Institute of Technology Madras, Chennai, INDIA} \\
\IEEEauthorrefmark{2}, {External Research, Global Research and
  Development, Ciena Corporation, Ottawa, CANADA} \\
{Corresponding Author: Krishna M. Sivalingam (e-mail:
  skrishnam@cse.iitm.ac.in); \\ Author emails: \{nv.shinde3@gmail.com,
  skrishnam@cse.iitm.ac.in, gshami@ciena.com\}}}

\maketitle

\begin{abstract}
    \acrfull{pdp} switches enable flexible packet processing but remain limited in scheduling capabilities, particularly for priority-based policies that require strict ordering of packets according to user-defined ranks. An earlier work, called \acrfull{pifo}, proposed a priority queue that enables ordering enqueued packets based on their priority. It provided an ideal abstraction for programmable packet scheduling; however, its requirement for line-rate packet sorting makes it impractical at high network speeds (100~Gbps and beyond). Existing accurate schedulers involve complex packet movement between the scheduling queues. Approximate schedulers such as \acrshort{sppifo}, \acrshort{aifo}, and \acrshort{rifo} reduce complexity but introduce priority inversions, thus degrading latency, fairness, and \Acrfull{fcts1} since they rely on \acrfull{fifo} queues or use \acrfull{aqm} as a substitute for scheduling. To address the scalability limitations of accurate schedulers while avoiding the correctness issues of approximate schedulers, this work proposes two packet queuing architectures, \acrfull{p3po} and \acrfull{ppqa}. 
     Both architectures decompose a large global priority queue into multiple smaller bounded-capacity priority queues that operate in parallel. \acrshort{p3po} uses a cascading set of priority queues with priority-based bounds to ensure that packets are inserted into the earliest queue capable of maintaining ordering. \acrshort{ppqa} distributes packets across multiple independent priority queues based on occupancy, using a de-multiplexer for balanced queuing and a multiplexer for globally selecting the highest-priority packet at dequeue. Both designs ensure bounded sorting complexity, preserve strict priority ordering by principle, and eliminate priority inversions.
    The architectures are implemented in the NetBench packet-level simulator and evaluated using empirical datacenter workloads (Web-search and Data-mining). Experiments are conducted on linear, leaf-spine, and star topologies using \acrfull{stfq} and pFabric-style ranking. The results show that \acrshort{p3po} achieves fair Bandwidth allocation, high Jain's Fairness Index, \acrshort{fcts1}, Per-packet delays, and Throughput comparable to the ideal \acrshort{pifo}, and performs better than the approximate schedulers. We show that \acrshort{p3po} and \acrshort{ppqa} have reduced sorting overhead compared to the ideal \acrshort{pifo}, thus reducing sorting latency. \acrshort{p3po} and \acrshort{ppqa} result in a 36.3\% and 89.3\% reduction in sorting overhead for small flows under different load levels. Analysis on datacenter topologies for high loads show that the proposed architectures perform better by 37.5~\% to 54.5~\% for smaller flow sizes and 21.4~\% to 38.8~\% improvement for large flows compared to approximate schedulers for \acrshort{stfq} scheduling. The per-packet delay in case of \acrshort{stfq} is lower by 67.9~\% for small flows to 70.7~\% for large flows. 

\end{abstract}

\begin{IEEEkeywords}
Approximate Schedulers; Packet Scheduling; Push-in First-out Queuing; Programmable Data Plane Switches. 
\end{IEEEkeywords}

\section{INTRODUCTION} \label{sect:introduction}
The emergence of programmable data plane switches  \cite{pdp-survey}
based on the Reconfigurable Match Action Table \cite{pisa-rmt}
architecture has enabled programming the forwarding behavior of data
plane switches using languages such as P4 \cite{p4}. However, only
certain functions related to the switch's behavior can be programmed,
such as new header definitions and packet processing. Traffic management
related functions such as Active Queue Management and Packet Scheduling
are still constrained to some extent since they can be configured but
not defined by the user. The Tofino\textsuperscript{TM} Native
architecture \cite{public-tofino} specifies that the Packet Scheduler in
the Traffic Manager of the switch can be configured with some Weighted
Fair Queuing \cite{wfq} mechanism. However, the implementation of the
scheduler is still a fixed-function model and cannot be modified by a
network operator.

At 100+~Gbps, switches have nanoseconds to process each packet,
requiring efficient handling of switching, inspection, and
scheduling. This limited time frame requires switches to perform
switching operations like packet parsing, processing, and applying
user-defined algorithms in addition to correctly scheduling packets. At
high line rates, scheduling inaccuracies can reduce fairness, cause
delays in latency-sensitive applications, and affect network
performance. 

The \acrfull{pifo} work \cite{pifo} introduced a programmable scheduler
abstraction using a priority queue, where packets are inserted by rank
(lower rank equals a higher priority) in their correct position, but are
dequeued only from the head of the queue. Thus, \acrshort{pifo} behaves
like a priority queue, where enqueued packets are ordered by their
ranks. A packet to be queued is always placed in its intended position
among the queued packets. \acrshort{pifo} enforces the property that any
packet dequeued from the priority queue always has the lowest rank or
equivalently the highest priority. Enqueuing a packet in its correct
position in abstraction \acrshort{pifo} requires comparing the incoming
packet with the existing packets in the queue until its corresponding
position is found. Due to this, \acrshort{pifo} abstraction faces
scalability issues as it needs to sort packets at the
line-rate. Implementations based on \acrshort{pifo} abstraction
generally reduce sorting overhead by reducing the buffer size to sort
packets at line rate.  
    
To address the scalability limitations of \acrshort{pifo} and have a
large buffer size, approximate schedulers such as PCQ \cite{pcq}, AFQ
\cite{afq}, Gearbox \cite{gearbox}, \acrshort{sppifo} \cite{sppifo},
\acrshort{aifo} \cite{aifo} and \acrshort{rifo} \cite{rifo} were
introduced. These schedulers approximate the behavior of PIFO using
multiple \acrfull{fifo} queues or using admission control
strategies. \acrshort{sppifo} \cite{sppifo} uses dynamic bounds that
help segregate packets into multiple \acrshort{fifo} queues. Each
\acrshort{fifo} queue represents a priority level in the scheduler, with
the packets in the queue with the highest priority \acrshort{fifo} being
served first. \acrshort{aifo} \cite{aifo} and \acrshort{rifo}
\cite{rifo} use admission control strategies with a \acrshort{fifo}
queue that enqueues only the highest priority packets based on the
occupancy levels of the \acrshort{fifo} queue or the relative rank of
the incoming packet. Packets with ranks that do not cross a certain
threshold are discarded. As both of these techniques are an
approximation of the abstraction \acrshort{pifo}, these mechanisms
introduce scheduling errors known as packet inversions. Packet inversion
is a state in which a lower priority packet is scheduled to be queued
before a higher priority packet. These scheduling errors introduce
increased latency and \acrshort{fcts1}, throughput fluctuations, and
affect fairness in bandwidth allocation.  

    To address the issue of scalability in large \acrshort{pifo} and priority inversions \cite{sppifo}, \cite{aifo}, \cite{rifo} introduced in approximate schedulers, we propose two parallel \acrshort{pifo}-based architectures called \acrfull{p3po} and \acrfull{ppqa} for programmable schedulers which can be implemented in traffic manager of a \acrshort{pdp} switch. \acrshort{p3po} and \acrshort{ppqa} use multiple concurrent priority queues to sort packets in parallel, each handling part of the rank space, thus reducing overall sorting latency.
    The key novelty of the proposed architectures lies in the idea of decomposing a large \acrshort{pifo} queue into smaller \acrshort{pifo} queues, which together can scale to large sizes while maintaining inversion-free scheduling and bounded sorting latency. By using smaller \acrshort{pifo} sub-queues, packet inversions are removed by design. \acrshort{p3po} also integrates overflow and under-utilization handling to maintain consistent performance in varying load conditions. This  work's key \textbf{contributions are summarized below:}
    
        \begin{itemize}
            \item Development of two architectures that leverage multiple \acrshort{pifo} queues to enable parallel sorting of a subset of packets by each queue.
            \item Implementation of both architectures using a packet level simulator.
            \item Analysis of reduced sorting overhead and fair bandwidth allocation
            \item Analysis of flow metrics over empirical workloads, viz. Web-search \cite{web-search} and Data-mining \cite{data-mining} with different scheduling policies and network topologies over different levels of network load.
            \item Demonstration of differential behavior in case of scheduling policies with and without starvation.
        \end{itemize}

    The existing and proposed architectures have been implemented in the NetBench packet-level simulator and evaluated using empirical datacenter workloads (Web-search and Data-mining). Experiments are conducted on linear, leaf-spine, and star topologies using \acrfull{stfq} and pFabric-style ranking. The results show that \acrshort{p3po} achieves performs better than the existing approximate schedulers in terms of fair bandwidth allocation,  Jain's fairness index, \acrshort{fcts1}, per-packet delays, and throughput, and is also comparable to the ideal \acrshort{pifo} scheduler.  We show that \acrshort{p3po} and \acrshort{ppqa} have reduced sorting overhead compared to the ideal \acrshort{pifo}, thus reducing sorting latency. \acrshort{p3po} and \acrshort{ppqa} result in a 36.3\% and 89.3\% reduction in sorting overhead for small flows under different load levels. The analysis of datacenter topologies for high loads shows that the proposed architectures perform better by 37.5\% to 54.5\% for smaller flow sizes and 21.4\% to 38.8\% improvement for large flows compared to approximate schedulers for \acrshort{stfq} scheduling. The delay per-packet in case of \acrshort{stfq} is lower by 67.9\% for small flows to 70.7\% for large flows. In case of pFabric scheduling, we show how priority queuing behaves with scheduling policies with inherent starvation issues. The artifacts for this paper are published on Github \cite{p3po-impl}.

    

The main additions and changes with respect to our previous conference
publication (\cite{p3po}) are: (i) Section~\ref{sect:bg-rltd-works}
provides additional relevant background material; (ii)
Sections~\ref{sect:main-ppqa} provides the additional architecture
\acrshort{ppqa} and its workflow; (iii) Section~\ref{sect:impl-analysis}
provides a micro-benchmark analysis of the proposed architectures by
measuring number of packet comparison operations
(Section~\ref{sect:impl-analysis-pkt-comps}) and Jain's Fairness Index
(Section~\ref{sect:impl-analysis-jfi}); (iv)
Section~\ref{sect:perf-eval} provides the flow-level performance
comparison with additional scheduling techniques; (v) The original paper
was based on the data center topology; additionally
Section~\ref{sect:perf-eval-star} provides performance evaluation on
star topology and Section~\ref{sect:perf-eval-dm} provides
representative results for Datamining workload.

    The rest of the paper is organized as follows. Section~\ref{sect:bg-rltd-works} provides the relevant background and related work. Section~\ref{sect:main} describes the proposed queuing architectures \acrshort{p3po} and \acrshort{ppqa} along with their algorithms and provides examples for enqueue and dequeue scenarios. Section~\ref{sect:impl-analysis} presents the implementation details, simulation environment, and provides the micro-benchmark analysis on a linear topology. Section~\ref{sect:perf-eval} provides a flow level performance by measuring \acrshort{fcts}, Per-packet delay, and Throughput for varying load levels. Section~\ref{sect:conclusions} concludes the paper.

\section{BACKGROUND AND RELATED WORK} \label{sect:bg-rltd-works}
    This section presents the background material and preceding research.
    \subsection{PROGRAMMABLE DATA PLANES}
\begin{figure}[hbtp]
            \centering
            \includegraphics[width=\linewidth]{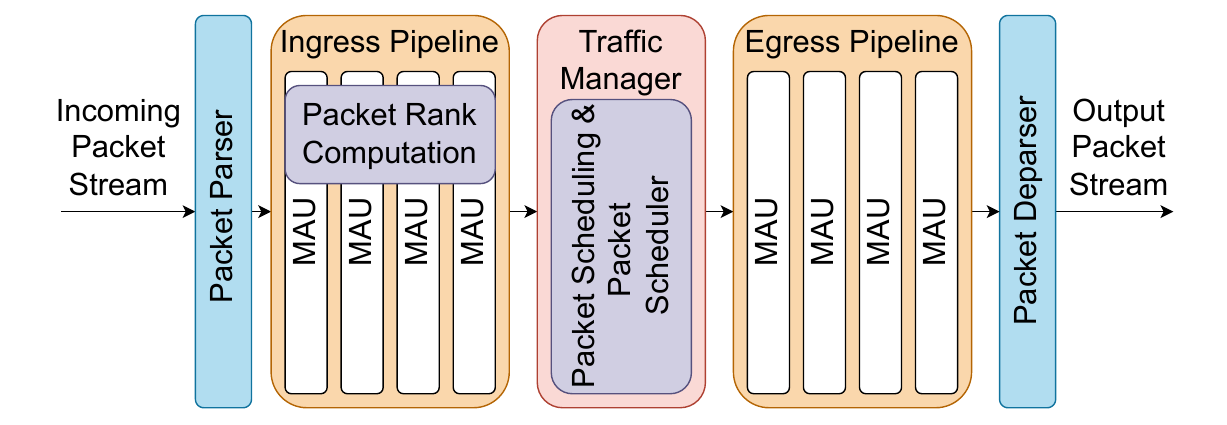}
            \caption{Programmable Data Plane with rank computation logic in Ingress pipeline and Packet scheduler.}
            \label{fig:pdp}
        \end{figure}

        Traditional network devices, such as routers and switches, operate on a fixed-function data plane, where packet processing logic is predefined by the device manufacturer. This fixed-function architecture limits adaptability, making it difficult to introduce new protocols, modify forwarding behavior, or implement custom network functions. A \acrfull{pdp} is a network forwarding plane that allows user-defined packet processing logic, enabling custom handling of network traffic beyond traditional IP and MAC-based forwarding. Unlike fixed-function switches, \acrshort{pdps} support custom forwarding, proprietary protocols, in-network computing, packet inspection and modification which could be implemented by a network operator. A simple data plane architecture is shown in Fig.~\ref{fig:pdp}. The traffic manager in current generation \acrshort{pdp} switches like Intel Tofino\textsuperscript{TM} implements a \acrshort{fifo} queue buffer where packet scheduling is performed which limits flexible packet scheduling.

\blueblock{
        \acrshort{pdps} decouple the network control plane from the data plane, allowing centralized or distributed controllers to install forwarding rules without modifying the underlying hardware. \acrshort{pdps} enable advanced use cases such as fine-grained traffic engineering, real-time telemetry, in-network aggregation, load balancing, and security enforcement.
}

    \subsection{PACKET SCHEDULING AND PROGRAMMABLE SCHEDULERS}\label{sect:bg-rltd-works-pkt-sched}
        Packet scheduling determines the order of transmission of packets to meet latency, fairness, and \acrfull{qos} requirements. Traditional packet schedulers operate on fixed algorithms implemented in hardware, such as First-Come First-Serve based on \acrshort{fifo} queues, Weighted Fair Queuing~\cite{wfq}, and Deficit Round Robin, thus limiting flexibility. While these schedulers expose configuration capabilities by modifying certain parameters (e.g. weights or quantum sizes), the underlying scheduling logic remains fixed, significantly limiting their flexibility. This fixed-function design does not allow network operators to implement custom scheduling policies customized to specific data center traffic requirements \cite{data-mining}, \cite{web-search} or dynamically changing traffic patterns. Fixed-function schedulers thus restrain fine-grained control over packet-level resource allocation.

        Programmable packet schedulers \cite{prog-sched}, \cite{ups} overcome these limitations by allowing users to define and modify scheduling policies dynamically. In programmable data planes, packets are generally assigned a rank (or priority) that encodes the desired scheduling order according to a user-defined policy. This rank of a packet can be computed within the network by network devices based on packet or flow characteristics (e.g. WFQ \cite{wfq}), device state, queue occupancy, or it can be explicitly set at end hosts, as demonstrated by pFabric~\cite{pfabric}. The scheduler then servers the packets according to the assigned rank, generally prioritizing packets with lower rank values (i.e., having higher priority).

\blueblock{
        In a programmable data plane switch architecture, the rank computation is implemented in the ingress pipeline, where programmable match-action stages compute the packet's scheduling metadata. This metadata is carried along with the packet in the traffic manager, which maintains packet queue buffers (\acrshort{fifo}) and schedules packets according to their ranks. After scheduling, packets are forwarded through the egress pipeline for final processing and transmission.
}

    \subsection{PRIORITY QUEUING AND PACKET PRIORITY INVERSIONS}\label{sect:bg-rltd-works-priority-que}
        Priority queuing enhances packet scheduling by classifying packets into multiple priority levels, where packets with higher priority are transmitted before those with lower priority. Each packet is assigned a priority value or rank based on factors such as application requirements (e.g., small flow completion times), latency sensitivity, flow characteristics, or bandwidth allocation (e.g., different weights to each flow). By enforcing priority ordering, priority queuing enables networks to provide network guaranties and meet \acrfull{qos} objectives. 
     
        Packet priority inversion occurs when a lower-priority packet is scheduled or dequeued ahead of a higher-priority packet, violating the intended priority ordering. Thus in schedulers based on \acrfull{fifo} queues, packets are processed strictly in the order of their arrival, irrespective of their assigned priorities. As a result, once a lower priority packet is enqueued into a \acrshort{fifo} queue ahead of a higher-priority packet, the scheduler cannot preempt it, leading to a priority inversion. Maintaining correct priority ordering based on the computed or assigned ranks is essential to ensure fairness and meet the Quality of Service (QoS) guaranties. However, several approximate priority schedulers, such as \cite{aifo}, \cite{rifo}, \cite{sppifo}, rely on \acrshort{fifo} queues and thus trade priority scheduling for less complex solutions. These schedulers trade strict priority enforcement for simplicity and scalability, inherently permitting packet priority inversions due to their \acrshort{fifo}-based structure.    
\blueblock{

        Although such approximate approaches are relatively simple to implement, priority inversions \cite{sppifo} can cause lower-priority packets to be dequeued before latency-sensitive or deadline-critical packets, resulting in increased tail latency and degraded application performance. Figs.~\ref{fig:fifo-q} and~\ref{fig:pifo-q} illustrate a non-priority (\acrshort{fifo}) queue and a priority queue, respectively. In a non-priority queue, packets are dequeued strictly in the order of arrival, without considering their assigned priority or rank. As shown in Fig.~\ref{fig:fifo-q}, although packets with ranks 1 and 2 are present in the queue, a packet with rank 5 is dequeued first because the queue does not enforce priority ordering. De-queuing a higher-rank (lower-priority) packet (e.g., rank 5) before a lower-rank (higher-priority) packet (e.g., rank 1) results in a packet priority inversion, thereby violating the intended priority ordering. In contrast, priority queuing ensures that packets are dequeued according to their ranks, so that the packet with the lowest rank (i.e., the highest priority) is always selected for transmission before any other packets. By preserving strict ordering based on priority, priority queues prevent packet priority inversions.

        \begin{figure}[htbp]
        \centering
            \subfloat[Non-Priority Queuing\label{fig:fifo-q}]{\includegraphics[width=\linewidth]{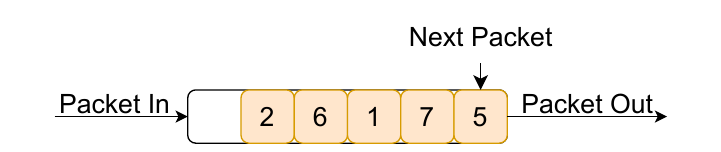}}\\
            
            \subfloat[Priority Queueing\label{fig:pifo-q}]{\includegraphics[width=\linewidth]{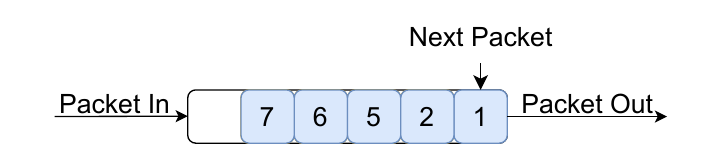}}
            \label{fig:fifo-pifo-qs}
        \end{figure}
}

\blueblock{

    \subsection{SCHEDULING POLICIES}\label{sect:bg-rltd-works-policies}
        pFabric~\cite{pfabric} is a transport-level scheduling policy designed for datacenter networks with the objective of minimizing flow completion times. It decouples rate control from flow scheduling and pushes scheduling decisions into the network, allowing switches to participate directly in packet scheduling. In pFabric, each packet is tagged with a priority that encodes the remaining size of its flow. Thus, packets belonging to flows with smaller remaining sizes are assigned higher priority than those of larger flows. Network switches then perform per-packet scheduling, transmitting higher-priority packets ahead of lower-priority ones. A known limitation of pFabric is packet reordering within a flow. Since the priority of a flow's packets decreases as the remaining flow size shrinks, packets that arrive later may be assigned a higher priority than earlier packets of the same flow. As a result, later-arriving packets can be dequeued before earlier ones, leading to packet reordering. Such reordering can cause starvation of earlier packets and impact QoS. Algorithm~\ref{alg:pfabric} presents the pseudo-code for the modified pFabric scheduling algorithm.

        \algrenewcommand\algorithmicindent{0.5em}
        \begin{algorithm}[htbp]
            \caption{pFabric Scheduling}\label{alg:pfabric}
            \begin{algorithmic}[1]
                \State \textbf{Global:}
                \Statex \quad $Q$ : Priority queue ordered by $(rank, enqueue\_id)$
                \Statex \quad $enqueue\_counter \leftarrow 0$

            \Procedure{Enqueue}{$P_{in}$}
                \State $f \leftarrow P_{in}.flow\_id$
                \State $r \leftarrow remaining\_flow\_size(f)$
                \State $P_{in}.rank \leftarrow r$
                \State $P_{in}.enqueue\_id \leftarrow enqueue\_counter$
                \State $enqueue\_counter \leftarrow enqueue\_counter + 1$
                \State Insert $P_{in}$ into $Q$
            \EndProcedure
        
            \Procedure{Dequeue}{}
                \If{$Q$ is not empty}
                    \State Remove and return packet with minimum $(rank, enqueue\_id)$ from $Q$
                \EndIf
            \EndProcedure
            \end{algorithmic}
        \end{algorithm}

        Max-min fairness \cite{wfq}, \cite{gps-multiple-node} is a resource allocation policy that distributes network capacity among competing flows in a fair manner. When multiple flows contend for a link, each flow is allocated an equal share of bandwidth until its demand is fully met. Any remaining capacity is then redistributed among the remaining flows. Weighted max-min fairness generalizes this concept by assigning weights to flows, allowing differentiated bandwidth allocation based on the relative importance of flows. Weighted max-min fairness is a generalization of this policy with weights assigned to each flow. Weighted Fair Queuing (WFQ) \cite{wfq} is a scheduling algorithm that implements weighted max-min fairness. In this work, we employ a simpler variant of WFQ known as Start Time Fair Queueing (STFQ) \cite{stfq}. STFQ reduces implementation complexity while maintaining fairness guarantees. 
        
        Algorithm~\ref{alg:stfq} provides the pseudo-code for the STFQ scheduling algorithm. In STFQ, each incoming packet is assigned a rank based on its flow's scheduling history and the system's virtual time. When a packet $P_{in}$ arrives, its flow identifier $f_{id}$ is determined. If the flow has been previously seen, the packet's start time is set to the maximum of the current virtual time and the flow's last recorded finish time; otherwise, it is initialized to the current system's virtual time. The flow's finish time is then updated based on the packet length and the flow's assigned weight $wt$. Packets with smaller ranks are dequeued first, enabling fair bandwidth sharing among flows while maintaining low computational overhead. 


        \algrenewcommand\algorithmicindent{0.5em}
        \begin{algorithm}[htbp]
            \caption{Start Time Fair Queuing (STFQ) Scheduling}\label{alg:stfq}
            \begin{algorithmic}[1]
            \State \textbf{Global:}
            \Statex \quad $Q$ : Priority queue ordered by $rank$
            \Statex \quad $virtual\_time \leftarrow 0$
            \Statex \quad $lastFinTime[\cdot]$ : Map from flow ID to finish time
            
            \Procedure{Enqueue}{$P_{in}$}
                \State $f_{id} \leftarrow flow(P_{in})$
                \State $wt \leftarrow weight(f_{id})$
            
                \If{$f_{id} \in lastFinTime$}
                    \State $P_{in}.start \leftarrow \max(virtual\_time,\; lastFinTime[f_{id}])$
                \Else
                    \State $P_{in}.start \leftarrow virtual\_time$
                \EndIf
            
                \State $P_{in}.finish \leftarrow P_{in}.start + \dfrac{P_{in}.length}{wt}$
                \State $lastFinTime[f_{id}] \leftarrow P_{in}.finish$
                \State $P_{in}.rank \leftarrow P_{in}.start$
                \State Insert $P_{in}$ into $Q$
            \EndProcedure
            
            \Procedure{Dequeue}{}
                \If{$Q$ is not empty}
                    \State $P \leftarrow$ remove packet with minimum $rank$ from $Q$
                    \State $virtual\_time \leftarrow P.rank$
                    \State \Return $P$
                \EndIf
            \EndProcedure
            \end{algorithmic}
        \end{algorithm}
}

    \subsection{RELATED WORK} \label{sect:bg-rltd-works-works}
        \acrfull{pifo} \cite{pifo} is a priority queuing abstraction that allows a packet to be inserted into a priority queue at any position based on its assigned rank while allowing dequeue to occur only at the head of the queue. Since \acrshort{pifo} is based on a priority queue, which enforces priority ordering, the dequeued packets are guaranteed to be the highest priority packets in the system. Thus, PIFO allows for inversion-free scheduling. \acrshort{pifo}'s model consists of scheduling transactions, which are micro compute units that compute a packet's rank based on a scheduling policy (e.g., STFQ \cite{stfq}, WFQ \cite{wfq}, pFabric \cite{pfabric}, etc) and a hierarchical scheduling structure where multiple queues are connected to form a tree for hierarchical scheduling. Every incoming packet in the scheduler must be inserted into the queue based on its priority, which involves sorting the queue elements. This requires comparing the incoming packet's rank with potentially all existing packets' priorities. This sorting needs to be performed at line rate to keep up with high-speed links (e.g., 0.8~ns per packet at 400~Gbps). PIFO thus struggles with scalability due to the need to sort packets at line-rate, which becomes difficult with large buffer sizes as line-rates exceed 100~Gbps.

\blueblock{
        To address scalability limitations of PIFO, approximate schedulers like SPPIFO \cite{sppifo}, AIFO \cite{aifo}, and RIFO \cite{rifo} were introduced. These schedulers use FIFO queues to approximate PIFO behavior using segregating packets into different queues, admission control, or both. SPPIFO \cite{sppifo} uses multiple FIFO queues to approximate PIFO by dynamically mapping packets to minimize scheduling errors (priority inversions). The packets are dequeued from the highest priority FIFO queue and then from subsequent priority FIFO queues. Each FIFO queue has an associated bound that determines the maximum priority of the packet that can be enqueued in that queue. SPPIFO implements a mapping algorithm that determines which packets are assigned to which queues based on the FIFO queue's bound and the packet's priority (rank). The algorithm also adapts these queue bounds when priority inversions occur, thus taking corrective measures to reduce future priority inversions. SPPIFO uses multiple queues, which are a limited resource in a network switch. 
      
        In contrast to \acrshort{sppifo}, \acrshort{aifo} \cite{aifo} re-frames the packet scheduling problem as an admission control problem. Specifically, it focuses on determining which packets should be admitted to the system so that priority inversions are reduced. \acrshort{aifo} uses a single \acrshort{fifo} queue; thus, once a packet is enqueued, its order cannot be changed according to future packet arrivals. \acrshort{aifo} maintains a sliding window of recently arrived packets to estimate the relative rank of new packets. It uses this rank to admit or drop a packet based on a dynamically calculated threshold, which ensures that low rank (high priority) packets are preferentially enqueued. The packets that do not pass the threshold check are dropped. Once enqueued, these packets are dequeued in strict order. 
        
        \acrshort{aifo} does not enforce rank-based ordering for packet dequeue, so packets are dequeued in the order of admission and not rank, leading to priority inversions. This may potentially increase latency for some packets even if \acrshort{aifo}'s queue state (set of enqueued packets at a particular time) matches the queue state of \acrshort{pifo}, since higher priority packets will be dequeued later. \acrshort{rifo} \cite{rifo} builds on the admission control strategy used by \acrshort{aifo}; however, by using fewer hardware resources on the \acrshort{pdp} switches. \acrshort{rifo} tracks the lowest and highest rank of packets in a given window of packets and uses a linear min-max normalization method to relatively rank packets based on recently observed packets in the window. Based on this relative rank, a packet is admitted based on the queue length. Additionally, since these techniques use an admission control strategy, packet drops are inherent to their design.
}

\greenblock{
        Sifter \cite{sifter} is a priority queuing architecture that uses multiple FIFO queues and a single small priority queue. The small priority queue stores packets with the highest priority (smallest ranks) and provides accurate sorting and always dequeues the packet with the smallest rank enqueued in the system. The set of \acrshort{fifo}s forms a Rotating Calendar Queue(RCQ), which stores packets with larger ranks, sorted with a coarser granularity. Each \acrshort{fifo} in the RCQ holds packets within a rank range. Sifter uses sift sorting to ensure that the highest priority packet is always available at the head of the priority queue. During the enqueue process, an incoming packet is either enqueued in the priority queue if its rank is less than the current least priority packet in the priority queue or in one of the \acrshort{fifo}s, depending on the rank range to which the rank belongs. A packet is always dequeued from the head of the small priority queue, which holds the smallest-rank packet.
}
    
        Since line speed sorting for larger queues is expensive in terms of time, \acrshort{p3po} and \acrshort{ppqa} use multiple smaller priority queues in order to reduce the number of sorting operations. Only the queue in which the packet is enqueued requires sorting. This ensures a bounded sorting latency. By adding additional priority queues to the system, P3PO can scale to large overall queue sizes while maintaining the inversion-free scheduling property.

        \begin{table}[htbp]
            \centering
            \caption{Comparison of existing approaches with proposed work}
            \label{table:comparison-table}
            \rowcolors{2}{blue!25}{white}
            \begin{tabulary}{\textwidth}{| L | C | C | C | C |} 
             \toprule
             \rowcolor{green!30}
             Work & Scheduling Order & No Inversions & Scalable & Scheduling Technique \\ [0.5ex] 
             \midrule \midrule
             PIFO \cite{pifo} & Accurate & \checkmark & \ding{55} & Priority \\ 
             \midrule
             SIFTER \cite{sifter} & Accurate & \checkmark & \checkmark & Mixed \\
             \midrule
             SP-PIFO \cite{sppifo} & Approximate & \ding{55} & \checkmark & Non-priority \\
             \midrule
             AIFO \cite{aifo} & Approximate & \ding{55} & \checkmark & Non-priority \\
             \midrule
             RIFO \cite{rifo} & Approximate & \ding{55} & \checkmark & Non-priority \\
             \midrule
             FIFO & No order & \ding{55} & \checkmark & Non-priority \\
             \midrule
             \rowcolor{yellow}
             P3PO & Accurate & \checkmark & \checkmark & Priority \\
             \midrule
             \rowcolor{yellow}
             PPIFO & Accurate & \checkmark & \checkmark & Priority \\ [1ex] 
             \bottomrule
            \end{tabulary}
        \end{table}

    \section{\textsc{Parallel Processing for Priority Ordering and Parallel PIFO Queuing Architectures}} \label{sect:main}
    
    This section describes the design and operation of the proposed parallel processing architectures.

\blueblock{
    \subsection{MOTIVATION} \label{sect:main-motiv}
        The primary limitation of traditional \acrshort{pifo}-based architectures is their requirement to perform packet sorting at line rate, which imposes significant hardware complexity and scalability constraints. To mitigate this overhead, approximate schedulers relax priority ordering requirements, but this introduces scheduling errors in the form of packet priority inversions.
        To address these challenges, we propose \acrshort{p3po} and \acrshort{ppqa}, two queuing architectures that decompose the global sorting operation into smaller, parallelizable units. The central idea behind both the architectures is to partition a logical large priority queue into multiple smaller sub-queues, each responsible for managing a subset of the overall traffic. These sub-queues operate independently and can be processed in parallel, thereby significantly reducing the per-queue sorting complexity.
        By parallelizing the sorting operation across multiple sub-queues, \acrshort{p3po} and \acrshort{ppqa} reduce sorting latency and improve throughput while preserving strict priority ordering. \acrshort{p3po} and \acrshort{ppqa} retain the inversion-free scheduling guaranties of \acrshort{pifo}-based schedulers by using priority queues without incurring the high line-rate sorting overhead, enabling scalable packet scheduling in programmable data planes.

    \subsection{DESIGN GOALS}  \label{sect:main-goals}
        Both \acrshort{p3po} and \acrshort{ppqa} aim to achieve the following key objectives:
        \begin{itemize}
            \item Bounded Sorting latency: \acrshort{p3po} and \acrshort{ppqa} partition the global priority queue into multiple sub-queues, each responsible for a disjoint subset of the rank space. By limiting the number of packets within each sub-queue, the sorting complexity and latency are bounded, enabling efficient line-rate operation. We show a reduction in sorting overhead by measuring the number of comparison operations performed when a packet is inserted into the scheduler.
            
            \item Priority Preservation: The enqueue logic in \acrshort{p3po} and \acrshort{ppqa} ensures that the highest-priority packet is always placed at the head of the sub-queues. This design resembles the behavior of a monolithic global priority queue, preserving strict priority ordering and preventing packet priority inversions.
            
            \item Sub-Queue Overflow Management: When a sub-queue reaches capacity, incoming packets are not immediately dropped. Instead, they are redirected to the next lower priority sub-queue. Packet drops occur only when the lowest-priority sub-queue overflows. This is applicable only to the \acrshort{p3po} architecture and not the \acrshort{ppqa} architecture. 
        \end{itemize}
}
    \subsection{PARALLEL PROCESSING FOR PRIORITY ORDERING ARCHITECTURE (P3PO)} \label{sect:main-p3po}
        
\begin{figure*}[htbp]
            \centering
            \includegraphics[width=\linewidth]{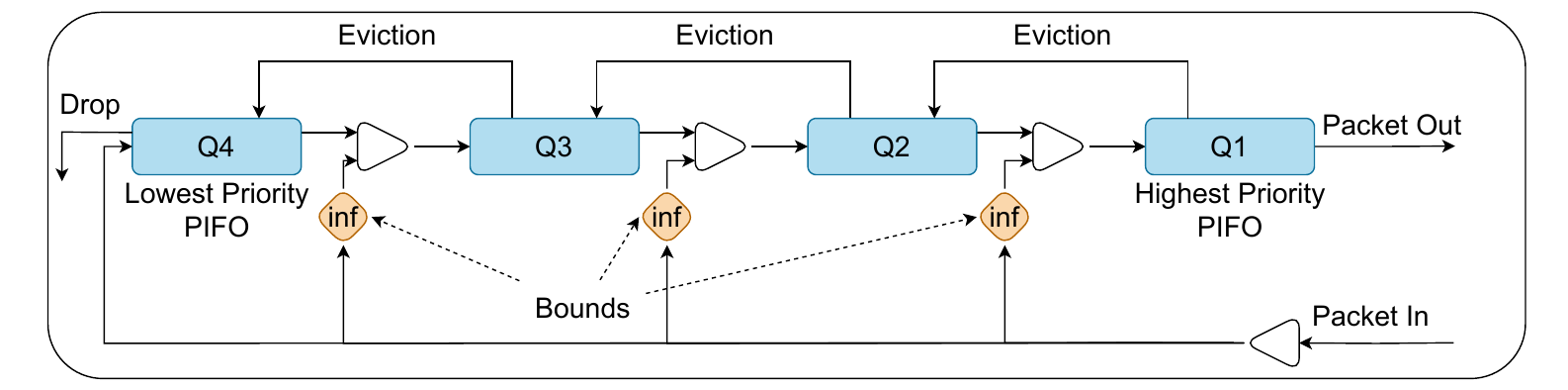}
            \caption{\acrfull{p3po}}
            \label{fig:p3po-arch}
        \end{figure*}

        The proposed priority queue \acrshort{p3po}, shown in Fig.~\ref{fig:p3po-arch} addresses PIFO's scalability limitation by decomposing a larger logical \acrshort{pifo} queue into $k$ smaller cascading priority queues of equal sizes $C$. Each queue $Q_i$ holds packets within a rank bound $b_i$, with bounds forming a non-decreasing sequence: $b_0\leq b_1 \leq \dots \leq b_{k-1}$. The packets are enqueued in the first queue where $r(p) \leq b_i$, where $r(p)$ is the rank (priority) of the packet. If the target queue is full, eviction cascades the lowest-priority packet to the next queue. If $Q_{k-1}$ is also full, the lowest-ranked packet is dropped from the system.
        When a packet arrives, its rank is compared sequentially against the bounds from $Q_0$ to $Q_{k-1}$. The packet is enqueued in the first queue, where its rank is lower than the queue's bound. If the target queue is full, an eviction cascade is triggered where the lowest-priority packet in the current queue is pushed in the next lower-priority queue, and so on. If the lowest-priority queue is full, the lowest-rank packet is dropped. This dropped packet has the lowest priority in the system.

    \subsection{P3PO WORKFLOW}
        In this section, we describe the working of \acrshort{p3po}. Each priority queue in the proposed architecture has an associated bound, which keeps track of the highest ranked (equivalently lowest priority) packet currently stored in that queue. The system follows the following steps for packet processing:

        Let $P = {p_1, p_2, \dots, p_n}$ be the incoming packets with ranks $r(p_i) \in \mathbb{Z}^+$ where a lower rank implies higher priority. Let $\mathbb{Q} = {Q_0, Q_1, \dots, Q_{k-1}}$ be $k$ priority queues and $B = {b_0, b_1, \dots, b_{k-1}}$ be the associated bounds. Incoming packets are assigned ranks based on the scheduling algorithm used in the network, e.g., pFabric \cite{pfabric} or the ranks computed in the switch's ingress pipeline according to the operator's requirements.
        
        \textbf{Initialization:} At the start of the scheduling operation, all queue bounds ${b_0, b_1, \dots, b_{k-1}}$ are set to the highest rank or the lowest priority that can be assigned to a packet in the network.

        \textbf{Packet Enqueue:} 
        The Algorithm~\ref{alg:P3PO-enq} presents the packet enqueue mechanism, which also handles the overflow mechanism. The rank of the incoming packet is extracted from the packet's metadata. The extracted packet rank (priority) is then compared against the bounds of each priority queue. A packet $p_{in}$ is inserted into the first queue $Q_j$ such that:
            $$r(p_{in}) \leq b_j \hspace{1cm} or \hspace{1cm} j=k-1$$
        where $r(p_{in})$ is rank of the packet. That is, the packet is inserted in the queue that has a bound higher than its rank or in the last queue. The bounds are compared from the highest priority queue to the lowest priority queue.

\begin{figure}[htbp]
        \centering
            \subfloat[State of the queues before packets with rank 8, 17, 19 are enqueued]{\includegraphics[width=\linewidth]{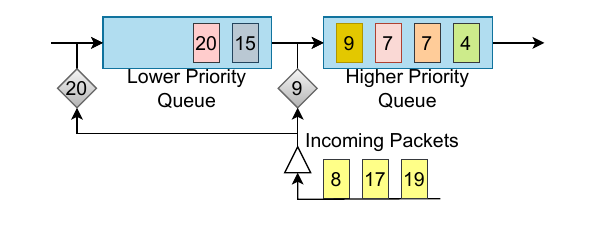}}\label{fig:p3po-ovf-a}\\[1ex]
            \subfloat[State of the queues after packets are enqueued and overflow handling is done]{\includegraphics[width=\linewidth]{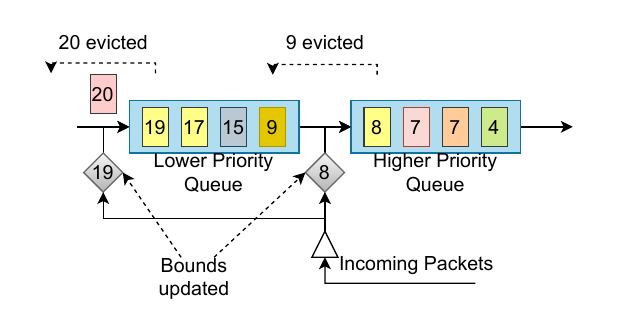}}\label{fig:p3po-ovf-b}
        \caption{Overflow handling}
        \label{fig:P3PO-overflow}
        \end{figure}

        \textbf{Overflow:} 
        If a $Q_i$ is full and receives a new packet, the lowest priority packet in $Q_i$ is evicted and inserted in $Q_{i+1}$. This process cascades until a queue with available buffer is found or the packet is dropped at $Q_{k-1}$ i.e. the last queue. Fig.~\ref{fig:P3PO-overflow} shows the process of handling an overflow situation. Fig.~{\ref{fig:P3PO-overflow}a.} shows the state where packets with ranks 8, 17, and 19 are to be enqueued. The higher priority queue ($Q_h$) is full with its bounds set to rank $9$, which is the last packet enqueued in it. The Fig.{~\ref{fig:P3PO-overflow}b.} shows the state of the queues after overflow handling is finished. Since the incoming packet rank is $8$, the packet with rank $9$ is evicted to the lower priority queue ($Q_l$). The bounds of $Q_h$ are updated to reflect the last packet in the queue, i.e. $8$. Similarly, packets with rank $17$ and $19$ are enqueued in $Q_l$ with its bound now updated to $19$ from $20$ since the packet with rank $20$ is evicted. The enqueue process performs an average of $k/2$ comparisons to select the appropriate queue when priorities are uniformly distributed and incurs an $O(C)$ cost for inserting the packet. In the event of an overflow, the complexity increases to $O(k \cdot C)$ to insert the packet in the correct queue.

        \textbf{Packet Dequeue:}
        Algorithm~\ref{alg:P3PO-deq} presents the packet dequeue mechanism, which handles under-utilization scenarios for all the queues except the least priority queue. This is because the least priority queue is the last in the architecture, thus there are no additional lower priority queues from where packets can be promoted. The scheduler always dequeues from $Q_0$ (or the highest non-empty queue). To make sure that packets are available to dequeue in the highest priority queue, when $Q_0$ is underutilized, packets are promoted from lower-priority queues using spillover thresholds in the.

\begin{figure}[htbp]
        \centering
            \subfloat[State of the queues where higher priority queue is underutilized]{\includegraphics[width=\linewidth]{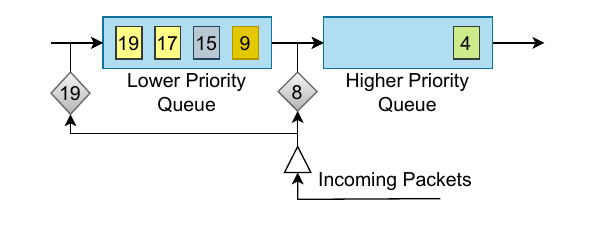}}\label{fig:p3po-und-a}\\[1ex]
            \subfloat[State of the queues after packets with rank 9, 15 are shifted to higher priority queue]{\includegraphics[width=\linewidth]{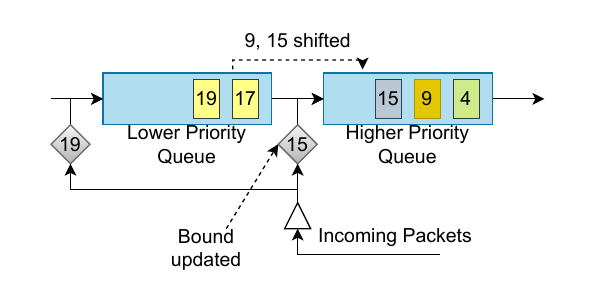}}\label{fig:p3po-und-b}
        \caption{Underflow handling}
        \label{fig:P3PO-underflow}
        \end{figure}

\greenblock{

        \begin{algorithm}[htbp]
            \caption{\acrshort{p3po} Packet Enqueue Process}\label{alg:P3PO-enq}
            \begin{algorithmic}[1]
                \Statex \textbf{Requires:}
                \Statex $P_{i} = \text{Incoming packet}$
                \Statex $Q_0, Q_1, \dots, Q_k-1 = \text{k Priority queues}$
                \Statex $b_0, b_1, \dots, b_k-1 = \text{k Queue bounds}$
                \Statex $C = \text{Queue Capacity}$
                
                \Statex
                
                \Procedure{Enqueue}{$P_i$}
                \State $Q=null$
                \State $r_i \gets P_i.priority$ \Comment{Rank}
                \For{$q=0$ to $k-1$}
                    \If{$r_i < b_q$ \text{OR} $q = k-1$}
                        \State insert $P_i$ in $Q_q$
                        \State $Q = i$
                        \State break
                    \EndIf
                \EndFor
                \If{$Q.len > C$}
                    \State \Call{Handle Overflow}{Q}
                \EndIf
                \EndProcedure   \Comment{Enqueued}
                
                \Statex     
                
                \Procedure{Handle Overflow}{$Q$}
                    \For{$q=Q$ to $k-1$}
                        \If{$Q_q.len > C$}
                            \State $P_t \gets Q_q.tail$     \Comment{Least Priority Packet}
                            \State Remove $P_t$ from $Q_q$
                            \State Update Bounds
                            \If{$q=k-1$}    \Comment{Last Queue}
                                \State Drop $P_t$
                            \Else
                            \State Insert $P_t$ in $Q_{q+1}$
                            \EndIf
                        \EndIf
                    \EndFor
                \EndProcedure
            \end{algorithmic}
        \end{algorithm}
}
        \textbf{Under-utilization:} 
        Each queue has a spillover threshold, which determines when it is considered underutilized. If $Q_i$'s occupancy drops below a threshold $\theta_l$ (e.g., 25\% of capacity), it pulls packets from $Q_{i+1}$ to maintain priority ordering and resource utilization. The bounds of each queue are adjusted according to the packet pulled into that queue. Fig.~{\ref{fig:P3PO-underflow}a.} shows the state where $Q_h$ is underutilized with its threshold set to 25\% of the queue capacity. Thus, packets with ranks $9$ and $15$ from $Q_l$ are pulled into $Q_h$, and its bound is updated to $15$, which is the rank of the last packet enqueued in the queue. During under-utilization handling, the packets are pulled until the queue occupancy reaches a higher threshold $\theta_h$ (e.g., 75\% of capacity). This ensures that a burst of higher-priority packets will not cause overflow handling with each enqueued packet. This mechanism helps maintain queue balance and ensures that packets enqueued in lower-priority queues are eventually dequeued and processed in the correct order. We noticed that in certain conditions, it happens that all the bounds are set to some ranks and the queues become empty as all the packets are dequeued. Thus, whenever a new packet arrives with a rank higher than the last priority queue's bound, the higher priority queues remain empty, with the packets getting enqueued in the least priority queue. This results in a large number of packet drops since the queues are severely underutilized. To address this issue, whenever queues become empty, all queue bounds ${b_0, b_1, \dots, b_{k-1}}$ are reinitialized to the maximum rank assigned in the network. The dequeue process occurs in $O(1)$ time, since the packet is dequeued from the head.

\greenblock{
        \begin{algorithm}[htbp]
            \caption{\acrshort{p3po} Packet Dequeue Process}\label{alg:P3PO-deq}
            \begin{algorithmic}[1]
                \Statex \textbf{Requires:}
                \Statex $Q_0, Q_1, \dots, Q_k-1 = \text{k Priority queues}$
                \Statex $b_0, b_1, \dots, b_k-1 = \text{k Queue bounds}$                
                \Statex $C = \text{Queue Capacity}$
                \Statex $\theta_l$ = Under Utilization Limit
                \Statex $\theta_h$ = Over Utilization Limit

                \Statex
                
                \Procedure{Dequeue}{}
                    \State \Call{Handle Underflow}{}
                    \State $P_o \gets Q_0.head$ 
                    \If{Total Length = $0$}
                        \State Reset queue bounds
                    \EndIf
                    \State \Return $P_o$
                \EndProcedure

                \Statex

                \Procedure{Handle Underflow}{}
                    \For{$q=0$ to $k-2$}
                        \If{$Q_q.len < \theta_l*C$}
                            \While{$Q_{q+1}.len \neq 0$ \&\& $Q_q.len < \theta_h*C$}
                                \State $P \gets Q_{q+1}.poll()$
                                \State $Q_q.insert(P)$
                                \State $b_q = P.priority$
                            \EndWhile
                        \EndIf
                    \EndFor
                \EndProcedure
            \end{algorithmic}
        \end{algorithm}
}

\blueblock{
    \subsection{PARALLEL PIFO QUEUING ARCHITECTURE (PPQA)}  \label{sect:main-ppqa}
        The proposed Parallel PIFO Queuing Architecture (PPQA)  architecture is illustrated in Fig.~\ref{fig:ppifo-arch}. It consists of multiple priority queues operating in parallel. On the ingress side, a demultiplexer compares the occupancy of all queues and directs each incoming packet to the queue with the lowest occupancy. If multiple queues have the same occupancy, the tie is resolved on the basis of the rank of the tail packets. On the egress side, a multiplexer block examines the head packets of all queues and selects the packet with the smallest rank for transmission. This design ensures balanced utilization of the available queues while maintaining strict priority-based packet scheduling at the output.    
        \begin{figure}[htbp]
            \centering
            \includegraphics[width=\linewidth]{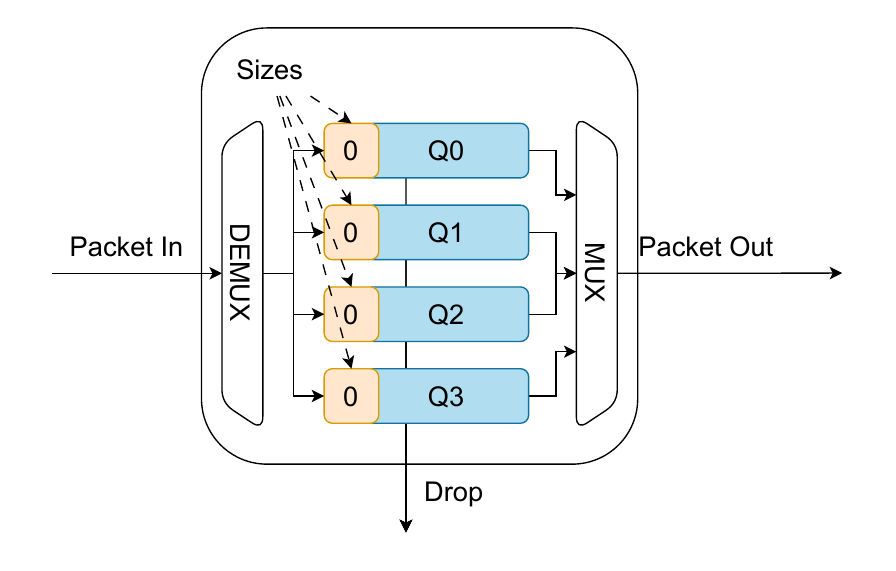}
            \caption{\acrfull{ppqa}}
            \label{fig:ppifo-arch}
        \end{figure}

    \subsection{PPQA WORKFLOW} \label{sect:main-ppqa-workflow}
    \acrshort{ppqa} is also built around multiple priority queues, with each queue handling a portion of the incoming traffic. Unlike \acrshort{p3po}, a packet enqueued in a particular queue will be dequeued without being moved to another queue. Each queue in this architecture has an associated size counter which tracks the current number of packets queued in the queue. These counters are updated whenever packets are enqueued or dequeued, and are used by the demultiplexer block to determine the queue in which to place the incoming packet. 
    
    There are two cases for packet enqueue: 1) A packet belongs to a flow that exists in the scheduler; in this case, the packet is enqueued in the same queue irrespective of the occupancy levels of other queues, and 2) The packet is the first packet of a new flow; in this case, the packet is enqueued in the queue with the lowest occupancy. This balances the load across the available queues and prevents overload in any single queue, as is the case with P3PO.  Packets that cannot be queued due to full capacity are dropped. On the output side, a multiplexer block examines the head packets of all the queues and selects the packet with the highest priority (lowest rank). The selected packet is then transmitted from the switch. This architecture ensures balanced distribution of incoming traffic across multiple queues while still enforcing priority-based scheduling at the output stage. 

    Fig.~\ref{fig:ppifo-enqueue-cases} shows two cases of packet enqueue in \acrshort{ppqa} architecture. As described earlier, in case of an incoming packet belonging to an existing flow as shown in Fig.~\ref{fig:ppifo-enqueue-case1}, the packet is enqueued in the same queue where the flow's earlier packets are enqueued. The incoming packet with rank $15$ is enqueued into the first \acrshort{pifo} queue as the packets of its flow are enqueued in that queue. In case of an incoming packet belong to a new flow as shown in Fig.~\ref{fig:ppifo-enqueue-case2}, the incoming packet with rank $1$ is enqueued into the queue with the lowest occupancy level which is the last \acrshort{pifo} queue. 

    \begin{figure}[htbp]
        \centering
            \subfloat[Case 1: Existing flow]{\includegraphics[width=\linewidth]{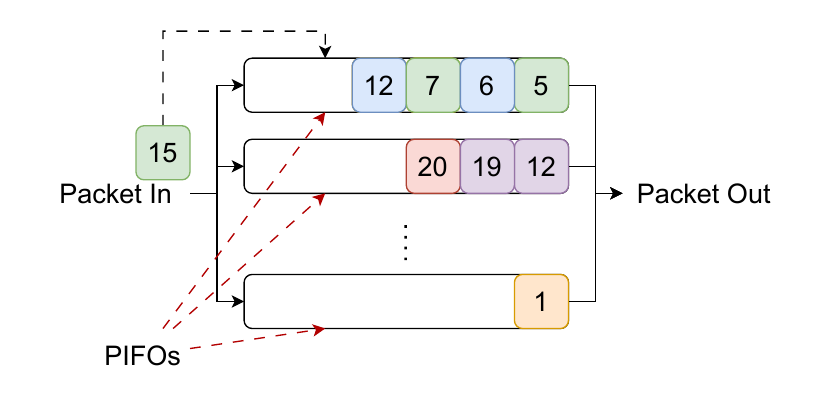}\label{fig:ppifo-enqueue-case1}}\\
            \subfloat[Case 2: New flow]{\includegraphics[width=\linewidth]{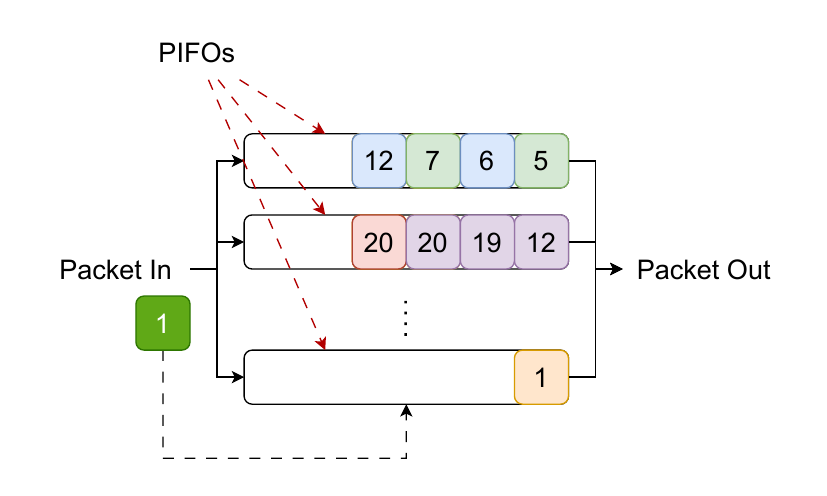}\label{fig:ppifo-enqueue-case2}}
        \caption{\acrshort{ppqa} enqueue cases}
        \label{fig:ppifo-enqueue-cases}
    \end{figure}
}
    \section{IMPLEMENTATION AND ANALYSIS} \label{sect:impl-analysis}
    This section presents the details of the implementation and presents a micro-benchmark analysis of \acrshort{p3po} and \acrshort{ppqa}.

\blueblock{
    \subsection{WORKLOAD DETAILS} \label{sect:impl-analysis-workloads}
        The analysis of \acrshort{p3po} and \acrshort{ppqa} architectures is performed on empirical datacenter flow distributions, Websearch \cite{web-search}, and Datamining \cite{data-mining}. The flows for both the distributions arrive according to a Poisson process, with the source and destination of each flow chosen in a uniformly random manner. The load in the topology depends on the arrival rates of these flows and corresponds to the average port utilization across all ports in the network. The Websearch workload has a mix of small and large flows from 10~KB to 20~MB, with 95\% bytes present in flows ranging from 1MB to 20~MB. The Datamining workload is skewed with 80\% of the generated flows lying below 10~KB. However, most of the bytes in the workload are from flows larger than 35~MB. We focus on Websearch workload in the analysis as the results show a similar trend for Datamining workload.
}
    \subsection{SIMULATION ENVIRONMENT} \label{sect:impl-analysis-environ}
        The proposed architectures were implemented in NetBench \cite{netbench}, a Java-based packet-level discrete event simulator. 
        The experiments were conducted on an Ubuntu 20.04 Linux machine with 128 GB RAM and a 32-core Intel Core i9-14900K processor. A new \acrshort{p3po} and \acrshort{ppqa} class extending NetBench's output port class was implemented to support the queuing mechanisms. For comparison, we used the existing \acrshort{sppifo} implementation \cite{sppifo-impl} along with the \acrshort{rifo} and \acrshort{aifo} implementations \cite{rifo-impl}, and recreated Sifter \cite{sifter} in NetBench based on the original paper, as the source code is not openly available. The set of \acrshort{fifo}s in Sifter is used as an ordinary group of FIFOs with equal granularity rather than as Rotating Calendar Queues (RCQ) \cite{calendar-queues}, \cite{pcq}. We implement an accurate mechanism of SIFTER without considering the speedup factor as mentioned in SIFTER \cite{sifter}.

\blueblock{
        \textbf{Schedulers configuration:}
        \acrshort{p3po} is configured with an under-utilization threshold of 80\%, which defines when higher priority queues pull packets from subsequent lower priority queues. Flows were generated using the empirical Web-search \cite{web-search} and Datamining \cite{data-mining} distributions. All schedulers are set up to have a large queue size of 1088 packets and are set to optimal values, as mentioned in their respective publications. \acrshort{p3po} and \acrshort{ppqa} each are split into 8 queues of equal size, while \acrshort{sppifo} has 8 \acrshort{fifo} queues of 136 packets each, and Sifter has one \acrshort{pifo} of size 64 packets and 8 \acrshort{fifo} queues of size 128 packets each. Sifter's \acrshort{pifo} is set to have a threshold of 50\% of its size, and during packet dequeue, it is refilled up to the threshold. \acrshort{aifo}, \acrshort{rifo}, and \acrshort{fifo} are configured with a First-In First-Out queue of 1088 packets in size.
}

        Flow metrics such as \acrfull{fcts}, Per-packet Delays and Throughput are analyzed for flows of different sizes for different levels of load under different scheduling policies, viz. \acrfull{stfq} and pFabric. We analyze \acrshort{p3po} and \acrshort{ppqa} with Datacenter topology, Linear topology, and Star topology to understand the behavior in different topology setups. The datacenter topology has 144 hosts, 9 leaf switches, and 4 spine switches, with leaf-spine links set to 400~Gbps and host-leaf links set to 100~Gbps. The linear topology consists of two hosts connected by a switch with 100~Gbps links. The star topology has 64 hosts and a single switch with links of 100~Gbps. Link delays are set to 1$\mu s$ in all topologies.

\begin{figure*}[hbtp]
            \centering
            \subfloat[10\% Load]{\includegraphics[width=0.33\linewidth]{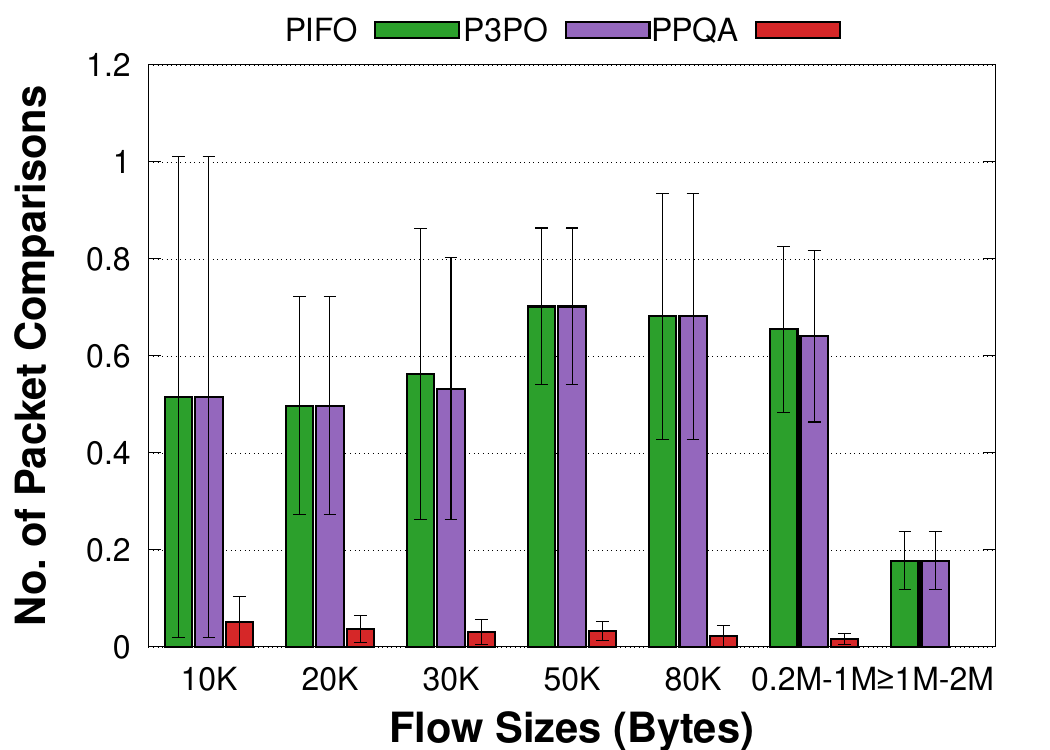}\label{packet-comps-wb-10}}
            \subfloat[50\% Load]{\includegraphics[width=0.33\linewidth]{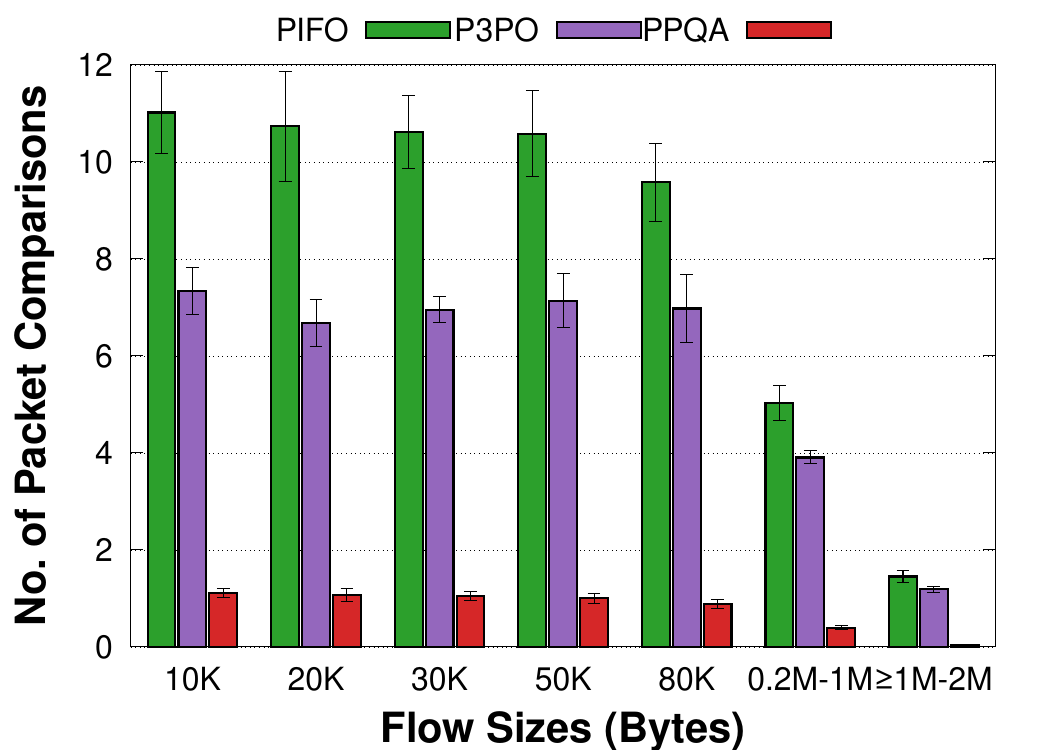}\label{packet-comps-wb-50}}
            \subfloat[90\% Load]{\includegraphics[width=0.33\linewidth]{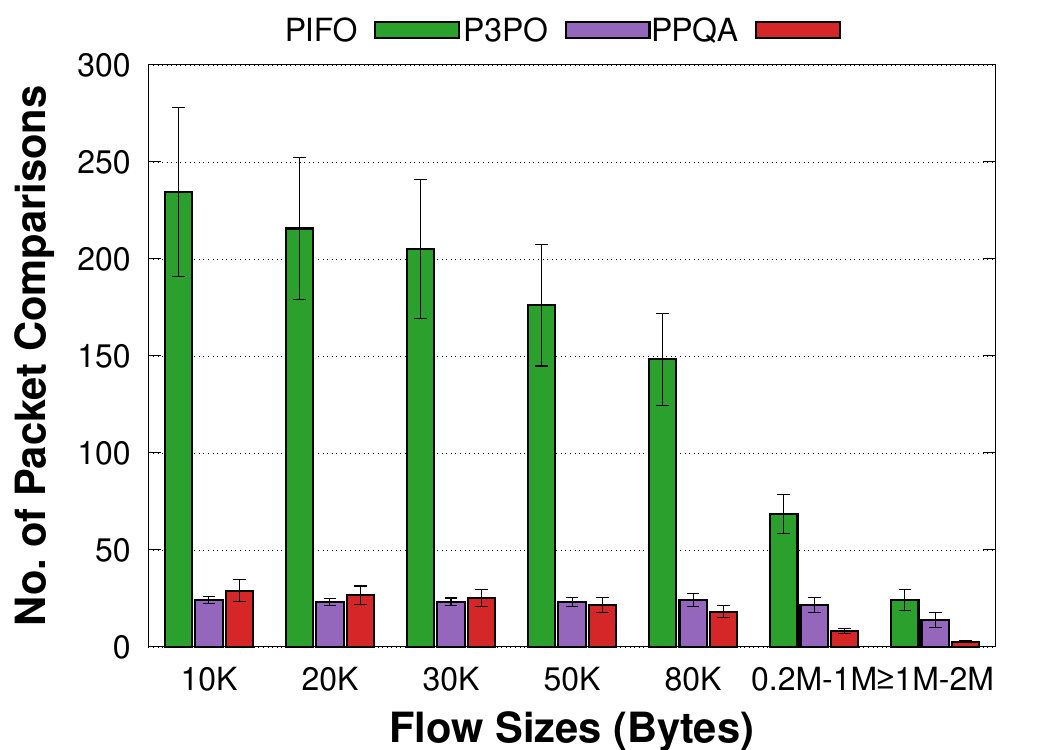}\label{packet-comps-wb-90}}
            \caption{Number of packet comparison operations for Websearch workload}
            \label{fig:packet-comps-wb}
        \end{figure*}

\begin{figure*}[hbtp]
            \centering
            \subfloat[10\% Load]{\includegraphics[width=0.33\linewidth]{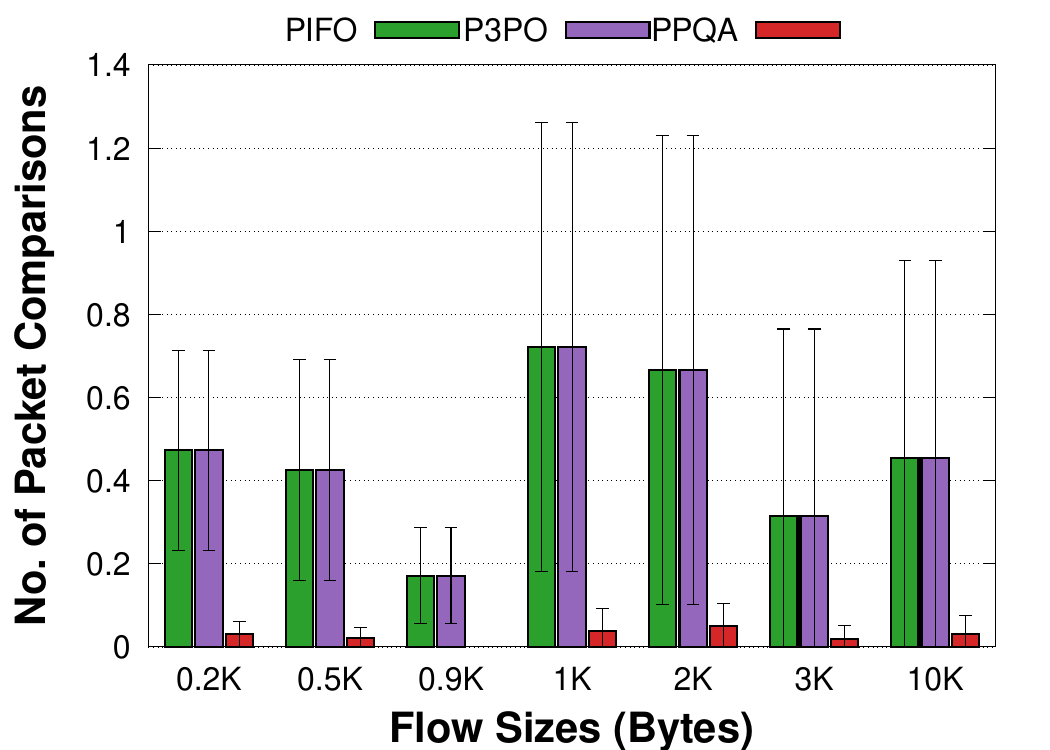}\label{packet-comps-dm-10}}
            \subfloat[50\% Load]{\includegraphics[width=0.33\linewidth]{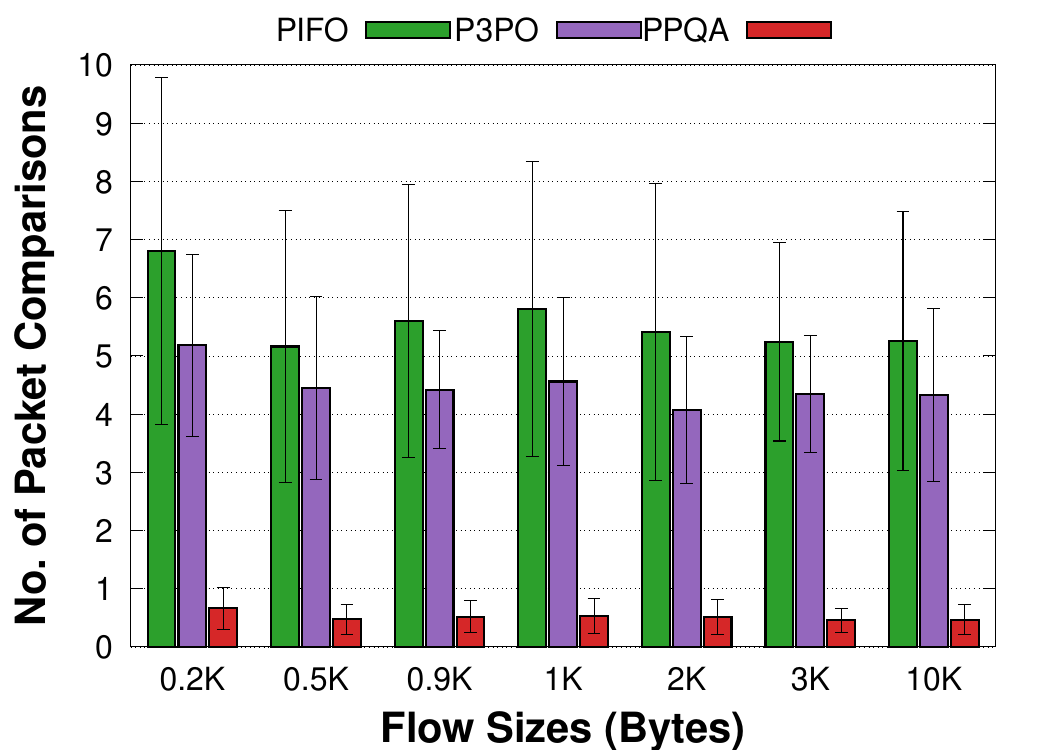}\label{packet-comps-dm-50}}
            \subfloat[90\% Load]{\includegraphics[width=0.33\linewidth]{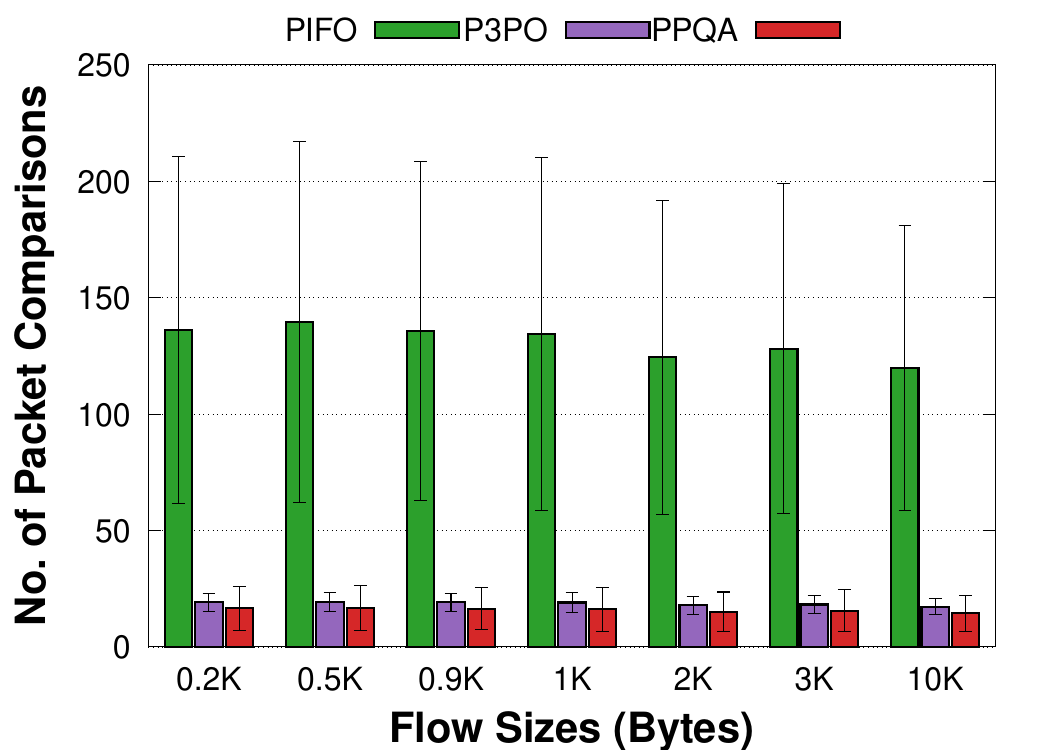}\label{packet-comps-dm-90}}
            \caption{Number of packet comparison operations for Datamining workload}
            \label{fig:packet-comps-dm}
        \end{figure*}

\blueblock{    
    \subsection{NUMBER OF PACKET COMPARISON OPERATIONS} \label{sect:impl-analysis-pkt-comps}
        To analyze the bounded sorting latency goal of \acrshort{p3po} and \acrshort{ppqa}, we run the queue mechanisms on a linear topology with two hosts connected by a switch with link speeds at 100~Gbps and the total queue size of 1088 packets with \acrshort{stfq} scheduling. We count the number of times a incoming packet is compared against already enqueued packets when being enqueued in the \acrshort{pifo}, \acrshort{p3po}, and \acrshort{ppqa} schedulers. Only these mechanisms are measured as they use the \acrshort{pifo} abstraction. Approximate schedulers and FIFO are not considered since they are based on the First-In First-Out mechanism, and thus, once a packet is enqueued, its position is not changed. 
        
        Simulation experiments were conducted for  loads of 10\%, 50\% and 90\%, resulting in differences in the number of packet comparison operations, which correlates directly with the queue occupancy levels. Fig.~\ref{fig:packet-comps-wb} shows the number of packet comparison operations at different load levels for Websearch workload. As seen in Fig.~\ref{packet-comps-wb-10}, for a load level of 10\%, the number of operations for a packet in \acrshort{pifo} and \acrshort{p3po} is similar. This is mainly because, since the load level is low, most packets are queued in the highest priority (first) \acrshort{pifo} queue in \acrshort{p3po}. Thus, any incoming packet is compared with the packets already queued until it is placed in its correct position. 
        
        In the case of \acrshort{ppqa}, since flows are separated into different queues based on occupancy levels, an incoming packet is compared against fewer packets, since the corresponding queue has fewer packets already enqueued. As the load increases, as seen in Fig.~\ref{packet-comps-wb-50} and Fig.~\ref{packet-comps-wb-90}, the enqueued packets are distributed into multiple queues in \acrshort{p3po} and \acrshort{ppqa}. 
        
        The percentage reduction in the number of packet comparisons observed from 50\% and 90\% increases with load. For 50\% load, \acrshort{p3po} has a 36.3\% reduction in the number of comparison operations for 10~KB flows and stays approximately consistent until 80~KB flow sizes. We observe that as the flow sizes increase, the percent reduction in the number of comparison operations decreases. 
        This is because smaller flows are frequent, and thus a packet from a recent small flow will have higher priority compared to a packet from an existing large flow. Thus, the incoming packet has to be compared with most enqueued packets to be placed closer to the head of the queue, based on its priority. This leads to higher number of operations for smaller flow sizes for \acrshort{pifo} scheduler. 
        
        For a load of 90\% as seen in Fig.~\ref{packet-comps-wb-90}, the percentage reduction in the number of packet comparisons is 89.3\% for flows of 10~KB, with a gradual decline as flow sizes increase. This can be attributed to the same reason as before. However, in this case, as the load is higher, packets are distributed into multiple priority queues in \acrshort{p3po}, leading to performance closer to \acrshort{ppqa}. In all cases, \acrshort{ppqa} performs significantly better, since flows are separated into multiple queues, and an incoming packet has to be compared with packets of the flow existing in the corresponding priority queue only. For 90\% load, the percentage reduction for \acrshort{ppqa} varies from 86.9\% for smaller flows to 85.7\% for larger flows.
        Similarly, Fig.~\ref{fig:packet-comps-dm} shows the number of operations for the Datamining workload. We observe similar trends compared to the Websearch workload.
}    

    \subsection{FAIRNESS OF BANDWIDTH ALLOCATION} \label{sect:impl-analysis-fairness}
            
        \begin{figure*}[htbp]
            \centering
            \includegraphics[width=\linewidth]{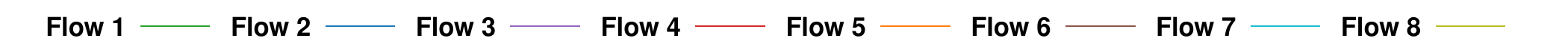}
            \includegraphics[width=\linewidth]{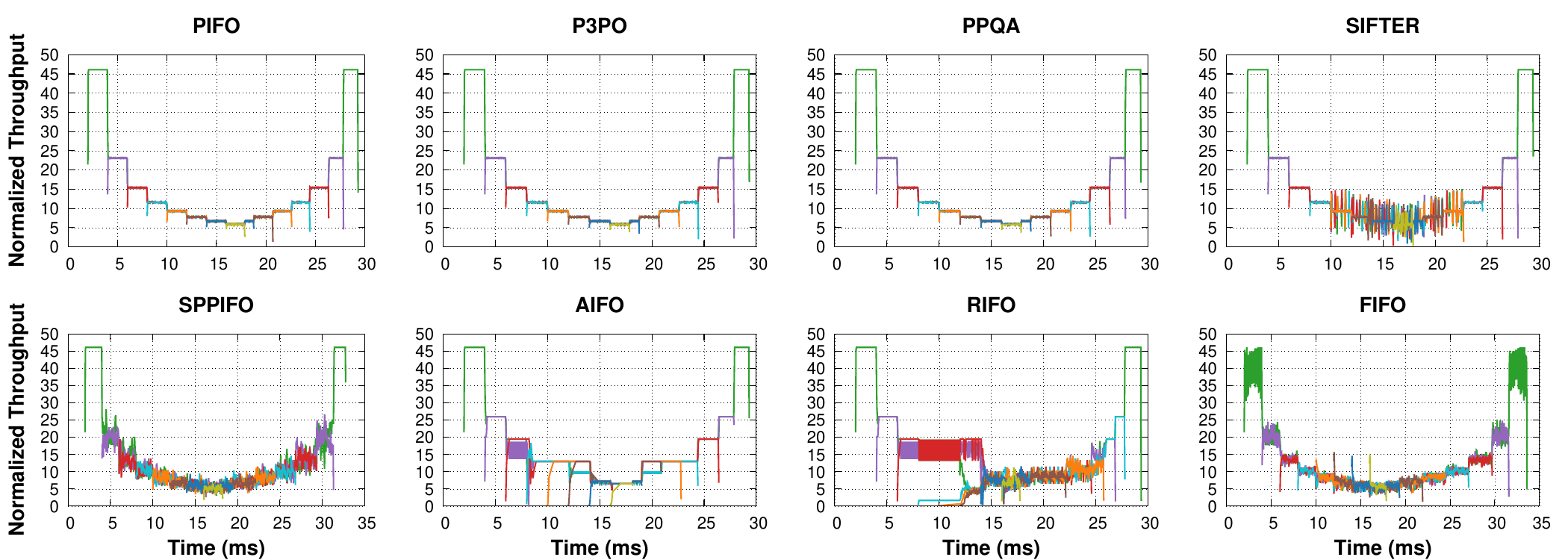}
            \caption{Fair bandwidth allocation with STFQ scheduling}
            \label{fig:flow-convergence}
        \end{figure*}

        We analyze the fairness of bandwidth allocation across different schedulers using the \acrshort{stfq} \cite{stfq} scheduling algorithm. \acrshort{p3po}, \acrshort{ppqa}, and other schedulers were simulated on a linear topology, which was discussed in an earlier section. Eight TCP flows start at the source host one after the other with an interval of 2~$ms$. The flows then finish in reverse order of their arrival at each other with an approximate interval of 2~$ms$. All flows are given equal weights. The throughput was measured every 40~$\mu s$, approximately ten times the round-trip time. Fig~\ref{fig:flow-convergence} presents the results of fair bandwidth allocation across eight flows for different schedulers. \acrshort{p3po}, with its multiple queues, behaves like a large \acrshort{pifo} queue. Thus, \acrshort{p3po} has its bandwidth allocated across flows similar to the \acrshort{pifo} queue. \acrshort{ppqa} follows a similar trend to \acrshort{p3po}, with the only difference being flows belong to different queues. Sifter shows fair bandwidth allocation up to the first 4 flows, but experiences fluctuations as additional flows are added. Since Sifter relies on multiple First-In First-Out queues, as the number of flows increases, packets are enqueued into different queues defined by each queue's rank range. 
        
        Sifter continuously sifts packets from the \acrshort{fifo} queues to its priority queue, to schedule the highest priority packet in the system. When a new flow is added, it is given a higher priority (lower rank) compared to existing flows. Hence, the ranks of existing flows increase (decrease in priority), and they are enqueued in the lower-priority \acrshort{fifo} queues. By the time the new flow is allocated bandwidth fairly, packets from existing flows that are enqueued in lower-priority queues are under served. As these packets need to be served, the packets from the flow that arrived later are given even lower priority. Thus, packets from earlier arrived flows are served better compared to packets from later arrived flows, leading to starvation. Thus, the loop of starvation of some flows and then fairly served is repeated, which causes fluctuations as the number of flows increases, which in turn leads to additional \acrshort{fifo}s being used. 
        
        In the case of approximate schedulers, \acrshort{sppifo} performs comparatively well, since it approximates \acrshort{pifo} behavior using multiple queues and does not rely on the admission control strategy used by \acrshort{aifo} and \acrshort{rifo}. Thus, \acrshort{sppifo} can separate priorities well between its \acrshort{fifo} queues and fairly prioritize flows with less allocated bandwidth, but it suffers from fluctuations since it relies on \acrshort{fifo} queues.
        
        We observe that \acrshort{aifo} and \acrshort{rifo} do not allocate bandwidth fairly. This can be attributed to their window based mechanisms and admission control strategy used. \acrshort{aifo}, considers the ranks of packets in a certain window and relatively ranks incoming packets to decide if they are admitted or not. Thus, when new flows are added to the system, the fair allocation fluctuates until the new flow is fairly given enough bandwidth, while packets from earlier flows are dropped since they do not cross its admission threshold. 
        \acrshort{rifo} uses a window based mechanism as well, however relies on the minimum and maximum packet ranks that are observed in the window. Thus when a new flow joins, the range of min rank and max rank increases significantly leading to fluctuations. 
        
        In addition, \acrshort{fifo} shows fair behavior in bandwidth allocation; however, this can be attributed to the inherent nature of \acrshort{fifo} to serve packets in order of their arrival. Thus, as additional flows are added, the number of packets in the system for each flow balances each other, giving the illusion of fair bandwidth allocation. If the flow weights were configured differently, \acrshort{fifo} will struggle to maintain the allocated bandwidth and demonstrate unfair behavior. The primary reason to make all flow weights equal is to measure Jain's Fairness Index as discussed in the subsequent section.

\blueblock{
    \subsection{JAIN'S FAIRNESS INDEX} \label{sect:impl-analysis-jfi}
        \begin{figure*}[htbp]
            \centering
            \includegraphics[width=\linewidth]{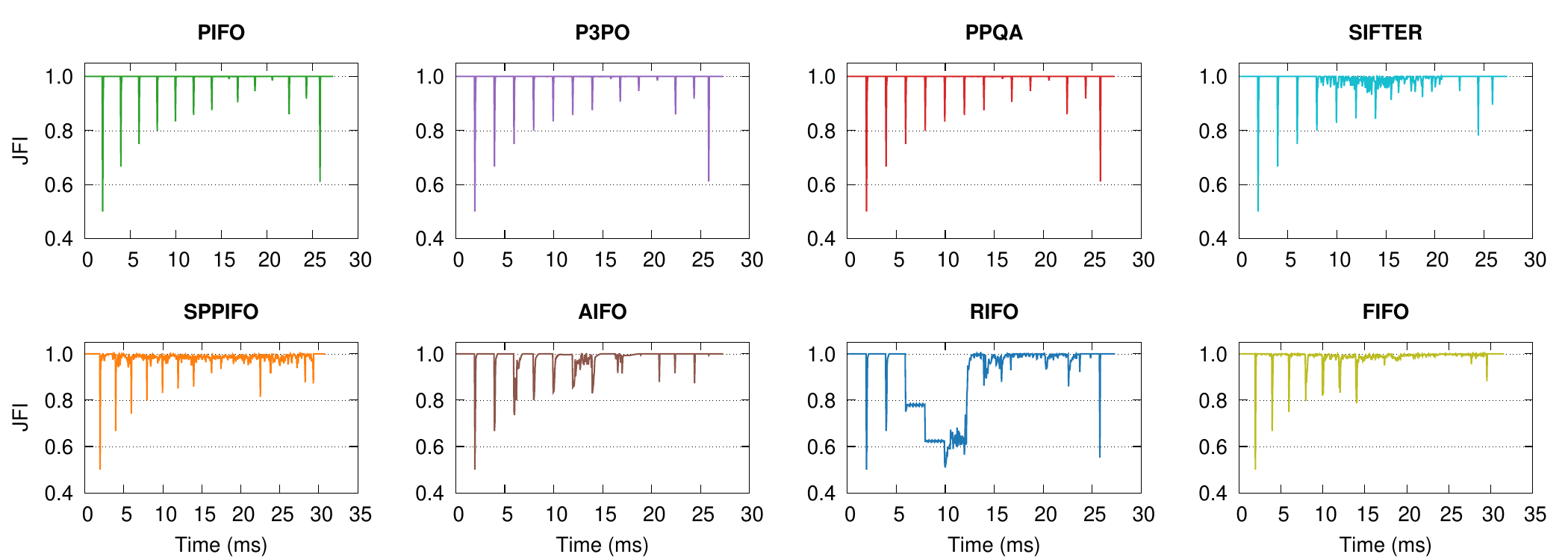}
            \caption{Jain's Fairness Index}
            \label{fig:jfi}
        \end{figure*}
        Jain's Fairness Index (JFI) \cite{jains-fairness} is a metric to measure how fairly or evenly a resource (link bandwidth) is shared between multiple flows. Each flow is given equal weight, and the following formula is used to measure JFI over a period of time. $$J(x_1,\ldots,x_n)=\frac{(\sum_{i=1}^{n} x_i)^2}{n\sum_{i=1}^{n} x_i^2}$$
        JFI has a range of $[\frac{1}{n}, 1]$, with a value of $\frac{1}{n}$ indicating worst-case behavior or unfairness, where a single flow is allocated the complete bandwidth. The value of $1$ indicates fair behavior with evenly distributed bandwidth between all flows. The Fig.~\ref{fig:jfi} shows Jain's Fairness Index across different schedulers over a period of $35ms$. The dips seen in the fairness across different schedulers correspond to a new flow being added to the system or marking flow completion. It can be observed that, as \acrshort{p3po} and \acrshort{ppqa} behave similarly to ideal \acrshort{pifo}, their fairness closely follows \acrshort{pifo}'s fairness. 
        
        Sifter, being an accurate scheduler, fairly allocates bandwidth to flows initially; however, the fairness varies when additional flows are added, for the same reasons as discussed in the previous section. Approximate schedulers struggle with fair bandwidth allocation, with fairness fluctuating when additional flows are added. \acrshort{sppifo} still maintains a fair allocation compared to \acrshort{aifo} and \acrshort{rifo}. As discussed earlier, AIFO and RIFO use a window based approach, which introduces tracking error. This leads to unfair behavior in bandwidth allocation between different flows. 
}

\section{FLOW LEVEL PERFORMANCE EVALUATION} \label{sect:perf-eval}
     This section presents a performance evaluation of the \acrshort{p3po} and \acrshort{ppqa} architectures when simulated on a data center topology with different workloads and load levels.

\purpleblock{
    \subsection{FLOW COMPLETION TIMES (FCTs)} \label{sect:perf-eval-fcts}
    Fig.~\ref{vary-load-dc-pfabric-fct} and Fig.~\ref{vary-load-dc-wfq-fct} demonstrate the flow completion times for a data center topology with 144 hosts, 9 leaves, and 4 spine switches for the Websearch workload. The server to host links are configured as 100~Gbps, and leaf to spine links are configured to 400~Gbps speeds. We simulate schedulers with the pFabric scheduling algorithm (Fig.~\ref{vary-load-dc-pfabric-fct}) and the \acrshort{stfq} scheduling algorithm (Fig.~\ref{vary-load-dc-wfq-fct}). As discussed in Section~\ref{sect:bg-rltd-works-policies}, pFabric assigns priorities to packets of a flow based on its remaining flow size. Thus, leading to starvation. We observe in Fig~\ref{vary-load-dc-pfabric-fct} that, with the pFabric scheduling algorithm, scheduling mechanisms that rely on priority ordering have higher flow completion times. This can be attributed to the packet starvation caused by the pFabric algorithm. It can be observed that for larger flows, the flow completion times are significantly lower in First-In First-Out queuing mechanisms compared to priority queuing mechanisms. This is because larger flows are starved more by pFabric, owing to its algorithm. 
    
    In contrast, when the STFQ algorithm is implemented, the flow completion times are lower for priority scheduling mechanisms, as there is no starvation. In both cases, \acrshort{p3po} and \acrshort{ppqa} perform in a way closely similar to the ideal \acrshort{pifo} queuing mechanism. Focusing on Fig.~\ref{vary-load-dc-wfq-fct}, as the load on the system increases, the priority queuing mechanisms perform better than the approximate mechanisms, as there are no packet inversions. \acrshort{p3po} and \acrshort{ppqa} closely follow the ideal \acrshort{pifo} scheduler. For lower loads, \acrshort{sppifo} has \acrlong{fcts} closer to \acrshort{pifo}, but as load increases, the number of packet inversions increase thus leading to larger flow completion times. Also, it can be observed that for \acrshort{sppifo}, the flow completion times are influenced by the flow size as well. 
    
    For flows above 80~KB, \acrshort{aifo} and \acrshort{rifo} perform slightly better than \acrshort{sppifo}. This is because, as load increases, \acrshort{sppifo} enqueues more packets in the scheduler, and \acrshort{aifo} and \acrshort{rifo} perform admission control and drop packets with relative ranks that do not cross a certain threshold. This leads to only some packets being enqueued in the \acrshort{aifo}, \acrshort{rifo} system, which are dequeued owing to the First-In First-Out policy. However, in the case of \acrshort{sppifo}, since packets are spread into multiple queues, packets in low-priority queues are starved, leading to higher FCTs. For 90~\% load, we observe an improvement of 37.5~\% to 54.5~\% for flows with 10~KB size, with the improvement being consistent across other small flows. For larger flows,  we observe an improvement of 21.4~\% to 38.8~\% compared to the approximate schemes. 
        \begin{figure*}[htbp]
            \centering
            \includegraphics[width=0.9\linewidth]{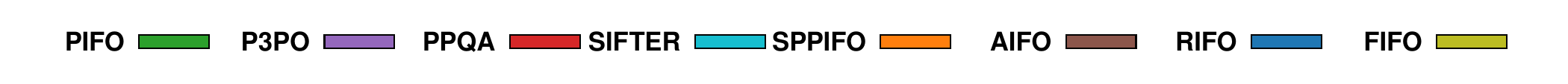}\\[-4ex]
            \subfloat[30\% load]{\includegraphics[width=0.32\linewidth]{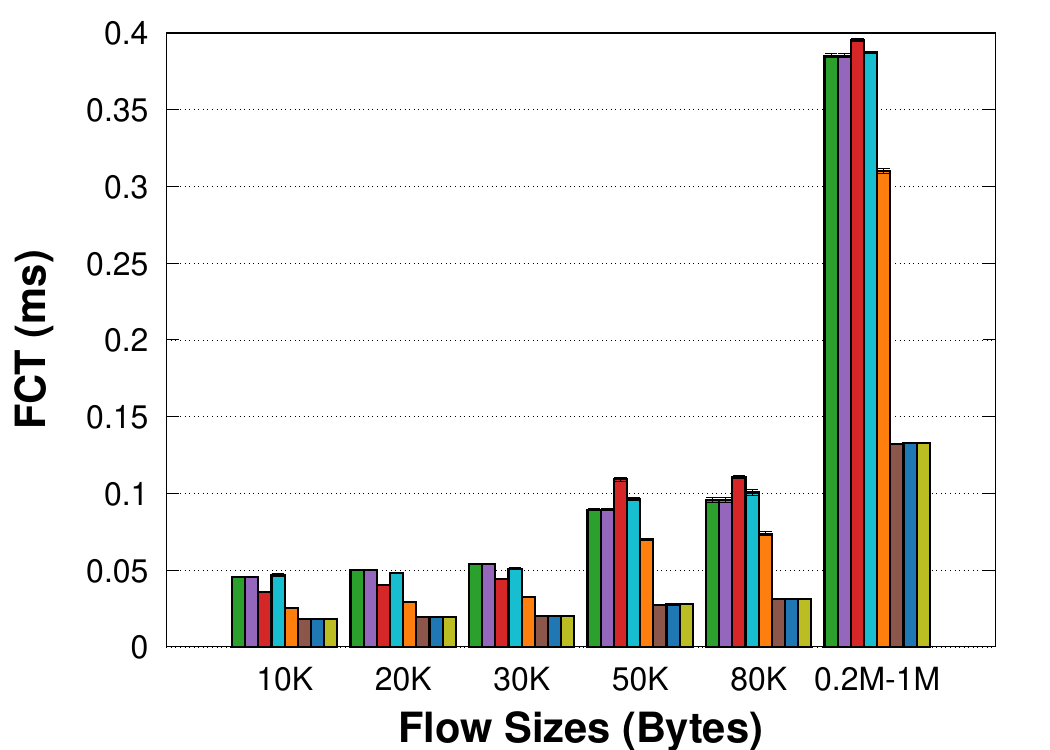}\label{vary-load-dc-pfabric-fct-30}}
            \subfloat[70\% load]{\includegraphics[width=0.32\linewidth]{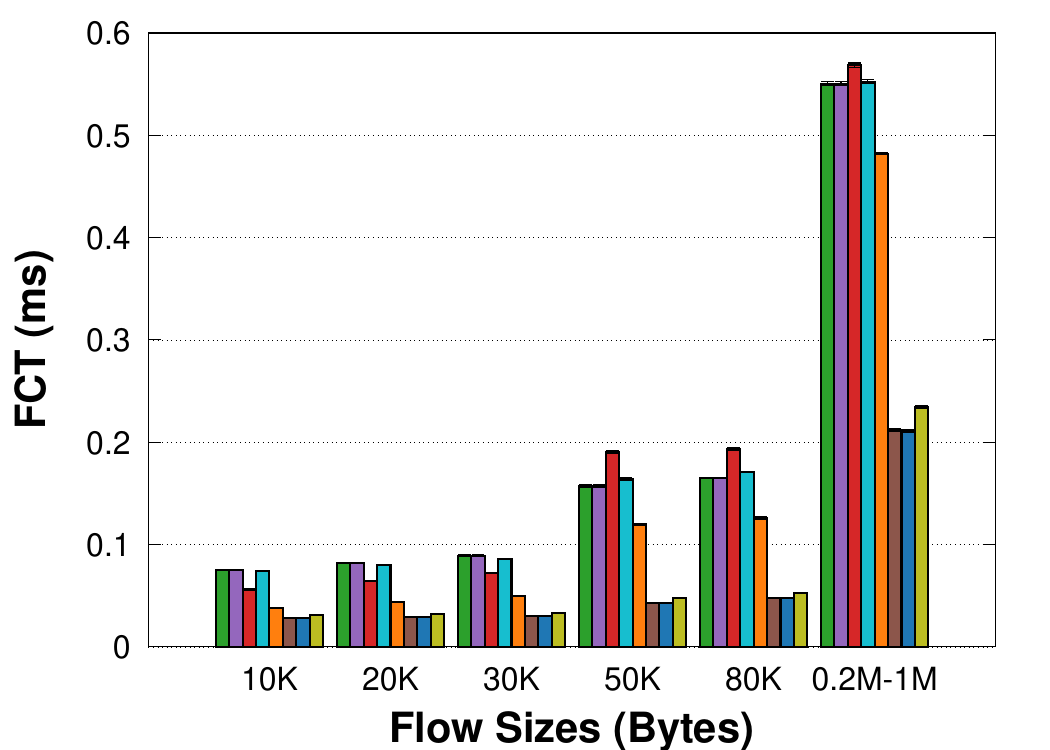}\label{vary-load-dc-pfabric-fct-70}}
            \subfloat[90\% load]{\includegraphics[width=0.32\linewidth]{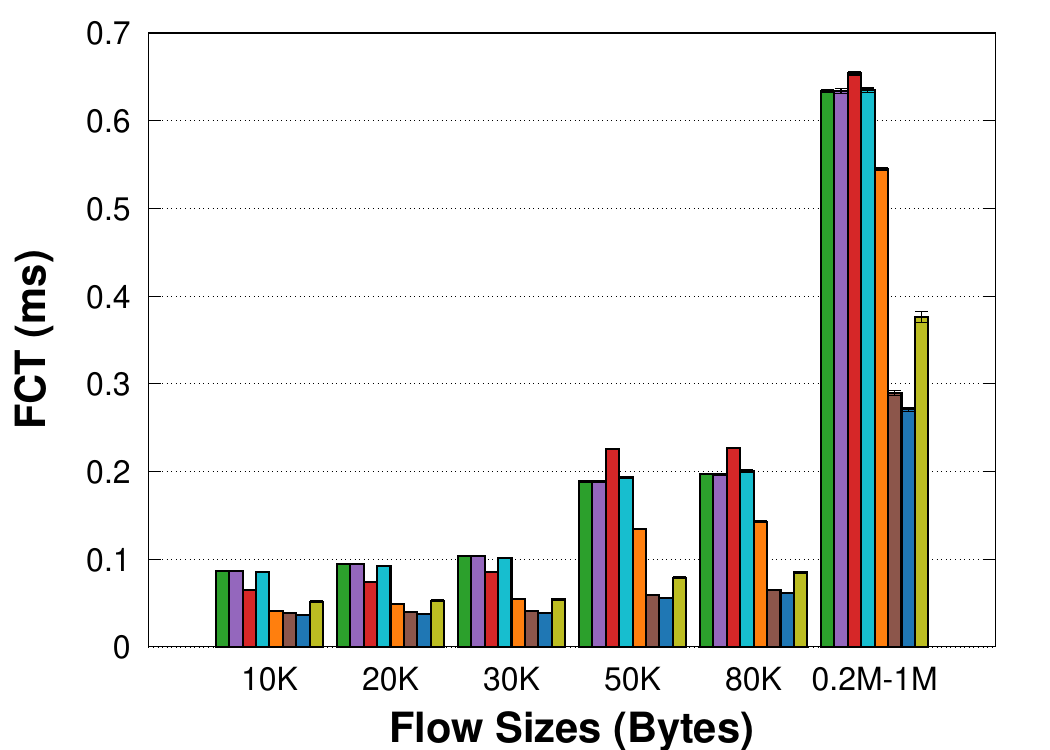}\label{vary-load-dc-pfabric-fct-90}}
            \caption{Flow Completion Times for Datacenter topology with pFabric scheduling}
            \label{vary-load-dc-pfabric-fct}
        \end{figure*}
    
        \begin{figure*}[htbp]
            \centering
            \includegraphics[width=0.9\linewidth]{journal-imgs/schedulers-legend-1rows.pdf}\\[-4ex]
            \subfloat[30\% load]{\includegraphics[width=0.32\linewidth]{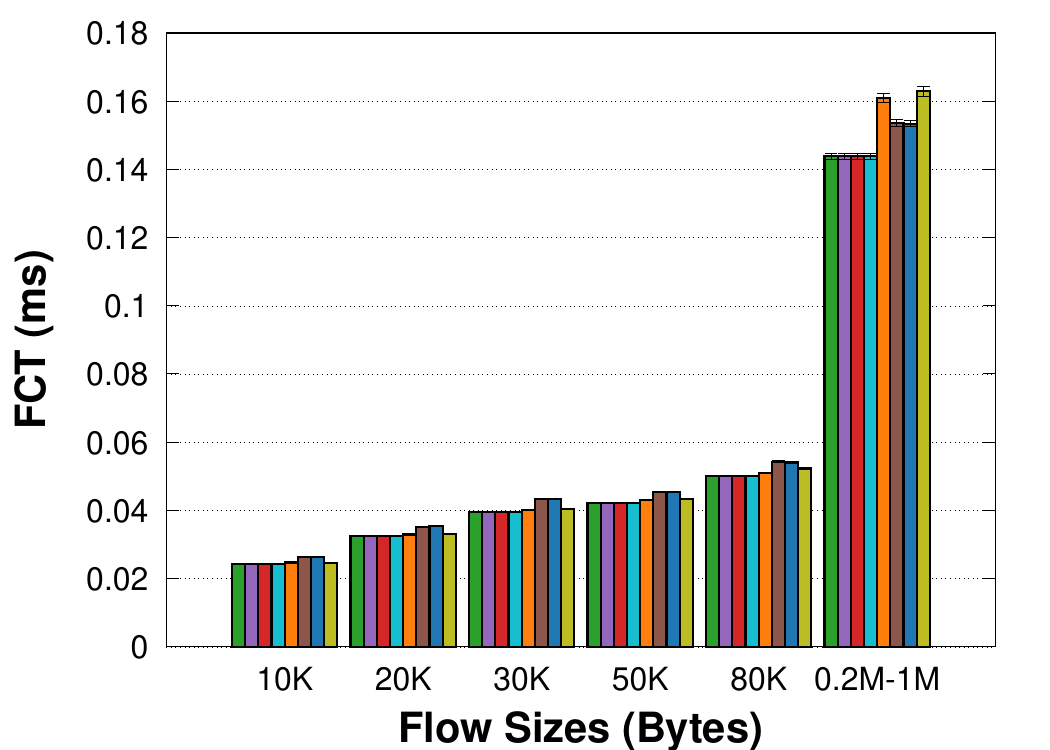}\label{vary-load-dc-wfq-fct-30}}
            \subfloat[70\% load]{\includegraphics[width=0.32\linewidth]{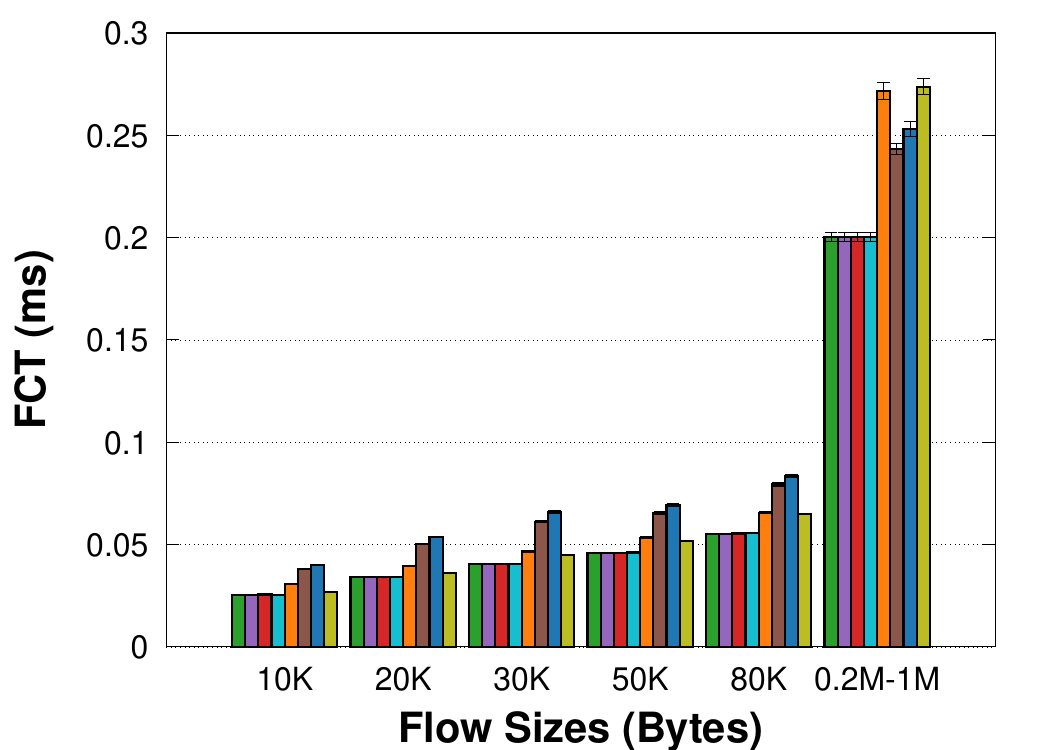}\label{vary-load-dc-wfq-fct-70}}
            \subfloat[90\% load]{\includegraphics[width=0.32\linewidth]{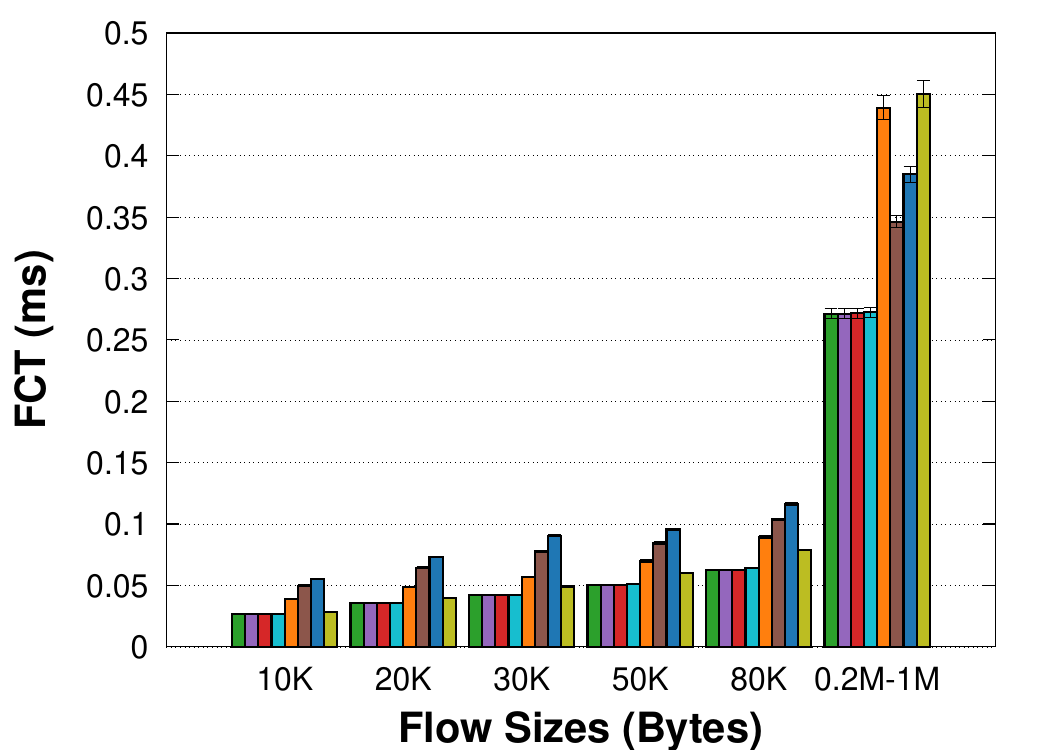}\label{vary-load-dc-wfq-fct-90}}
            \caption{Flow Completion Times for Datacenter topology with STFQ scheduling}
            \label{vary-load-dc-wfq-fct}
        \end{figure*}
}

\begin{figure*}[htbp]
            \centering
            \includegraphics[width=0.9\linewidth]{journal-imgs/schedulers-legend-1rows.pdf}\\[-4ex]
            \subfloat[30\% load]{\includegraphics[width=0.32\linewidth]{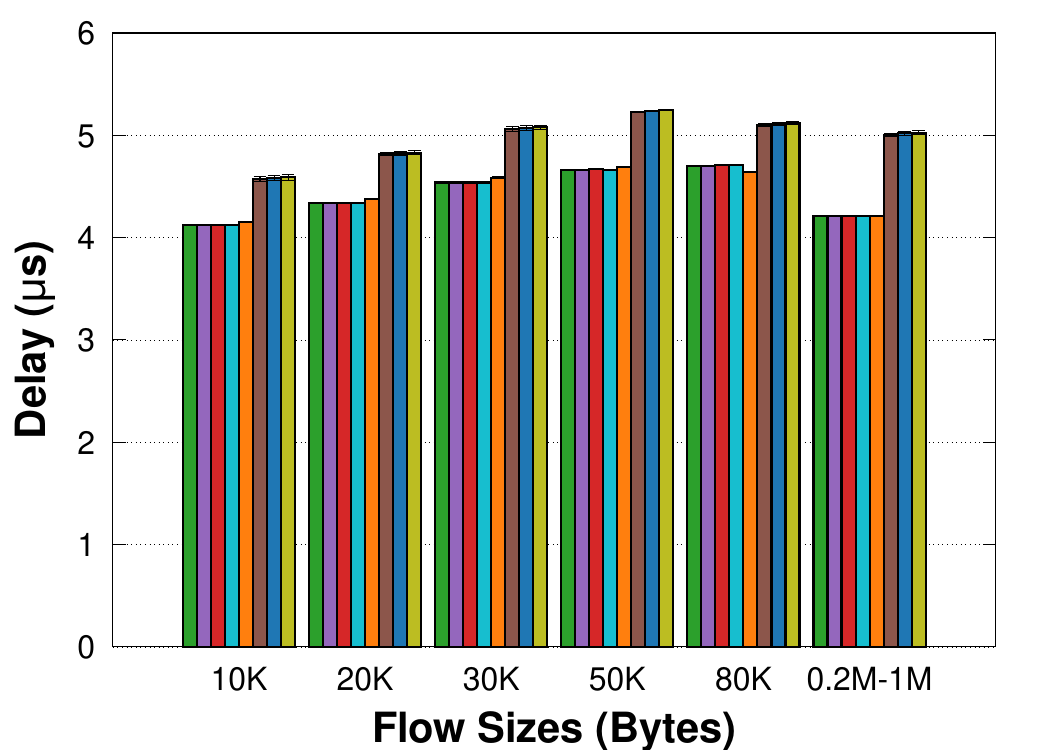}\label{vary-load-dc-pfabric-delay-30}}
            \subfloat[70\% load]{\includegraphics[width=0.32\linewidth]{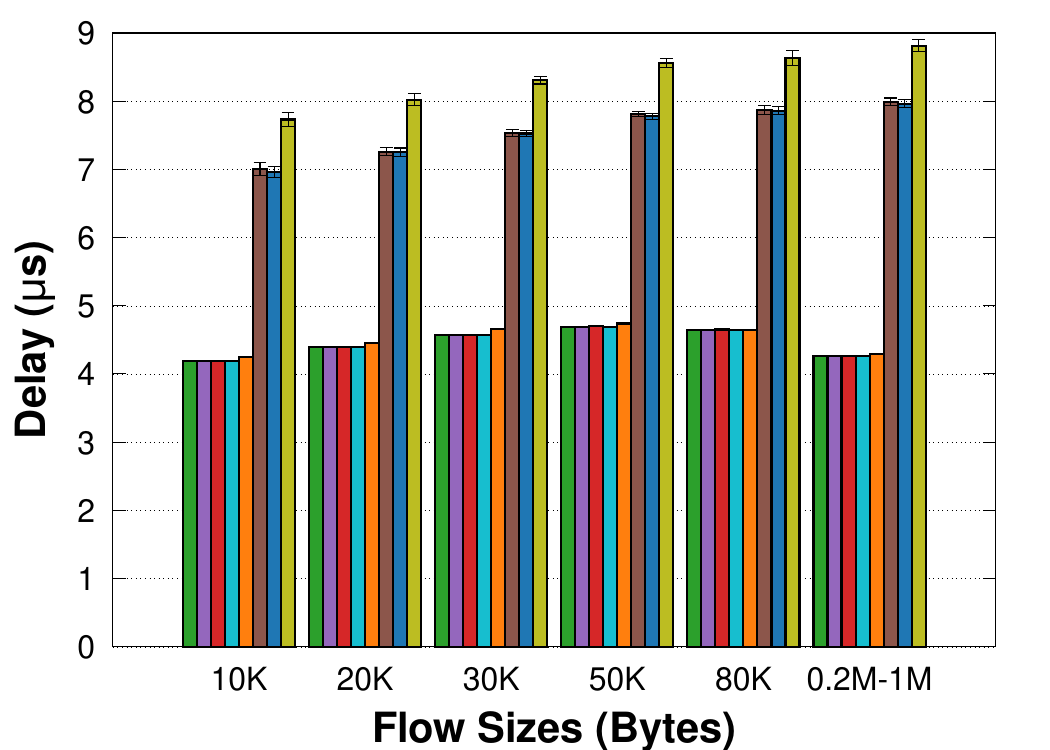}\label{vary-load-dc-pfabric-delay-70}}
            \subfloat[90\% load]{\includegraphics[width=0.32\linewidth]{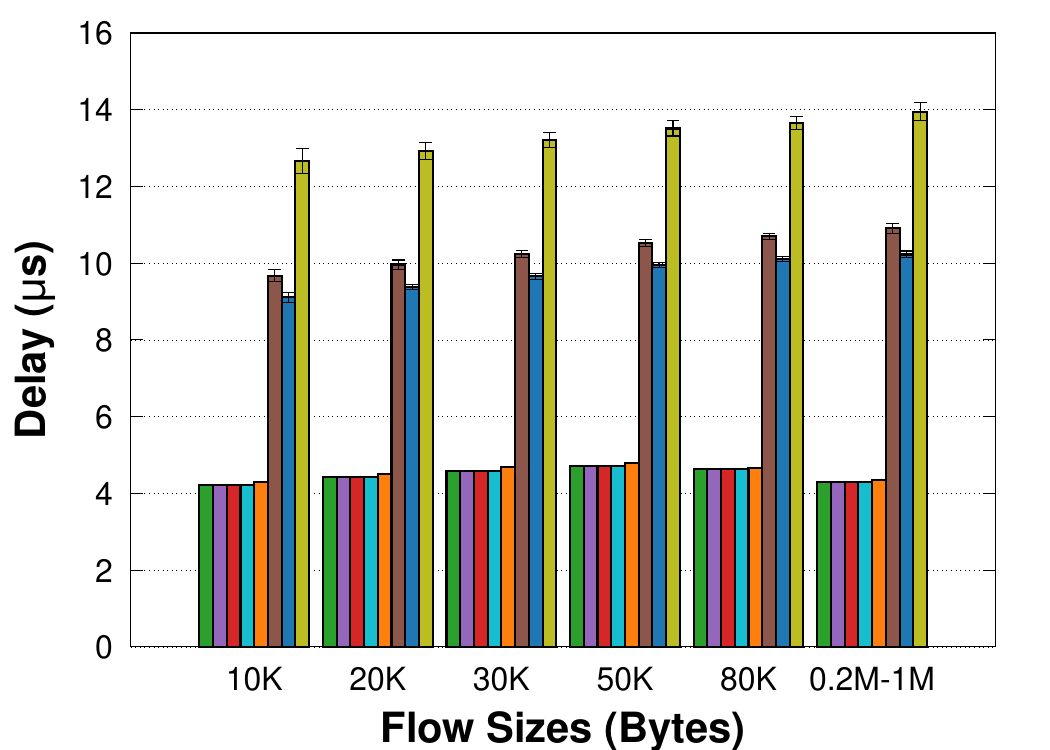}\label{vary-load-dc-pfabric-delay-90}}
            \caption{Per packet delay for Datacenter topology with pFabric scheduling}
            \label{vary-load-dc-pfabric-delay}
        \end{figure*}

\begin{figure*}[htbp]
            \centering
            \includegraphics[width=0.9\linewidth]{journal-imgs/schedulers-legend-1rows.pdf}\\[-4ex]
            \subfloat[30\% load]{\includegraphics[width=0.32\linewidth]{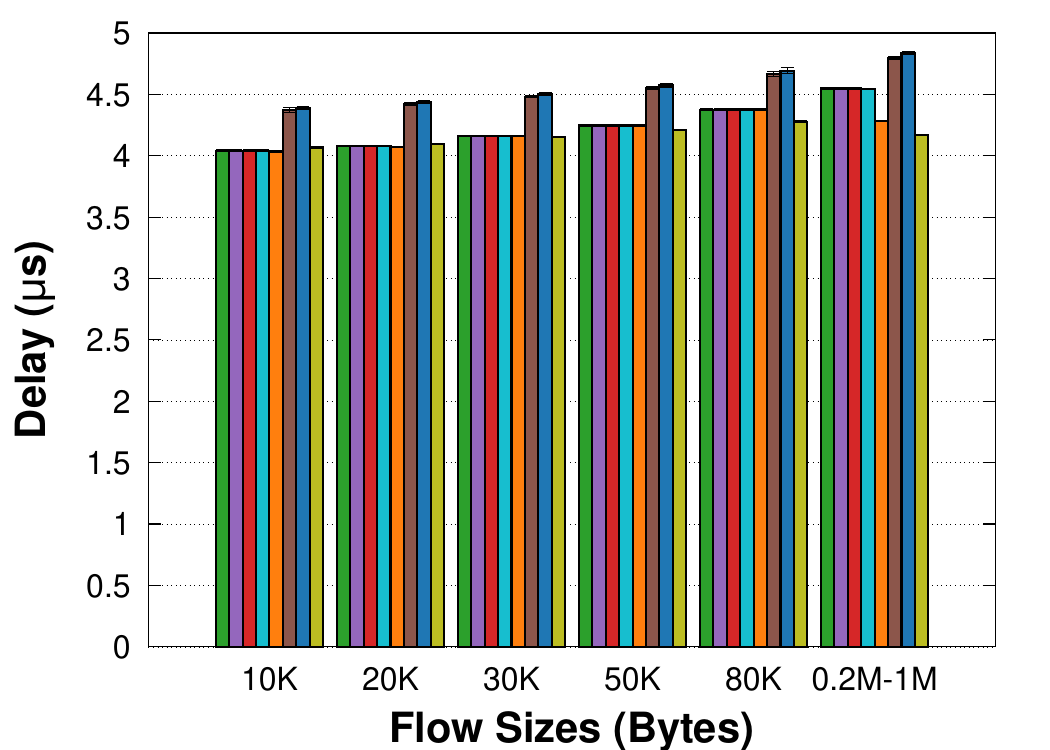}\label{vary-load-dc-wfq-delay-30}}
            \subfloat[70\% load]{\includegraphics[width=0.32\linewidth]{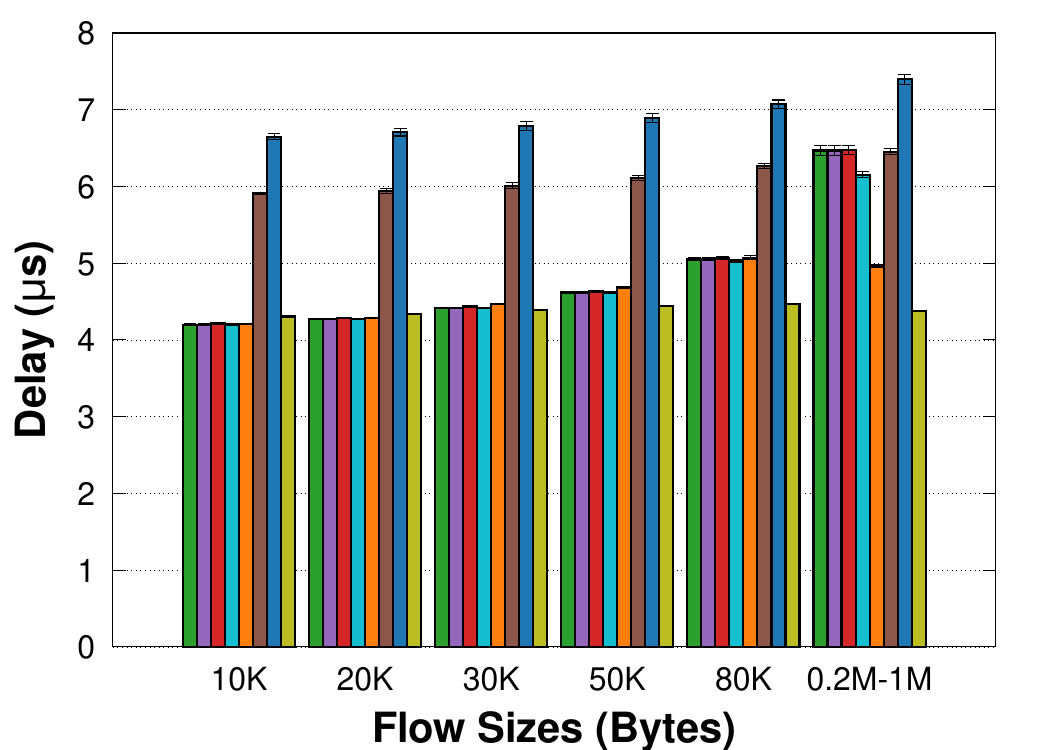}\label{vary-load-dc-wfq-delay-70}}
            \subfloat[90\% load]{\includegraphics[width=0.32\linewidth]{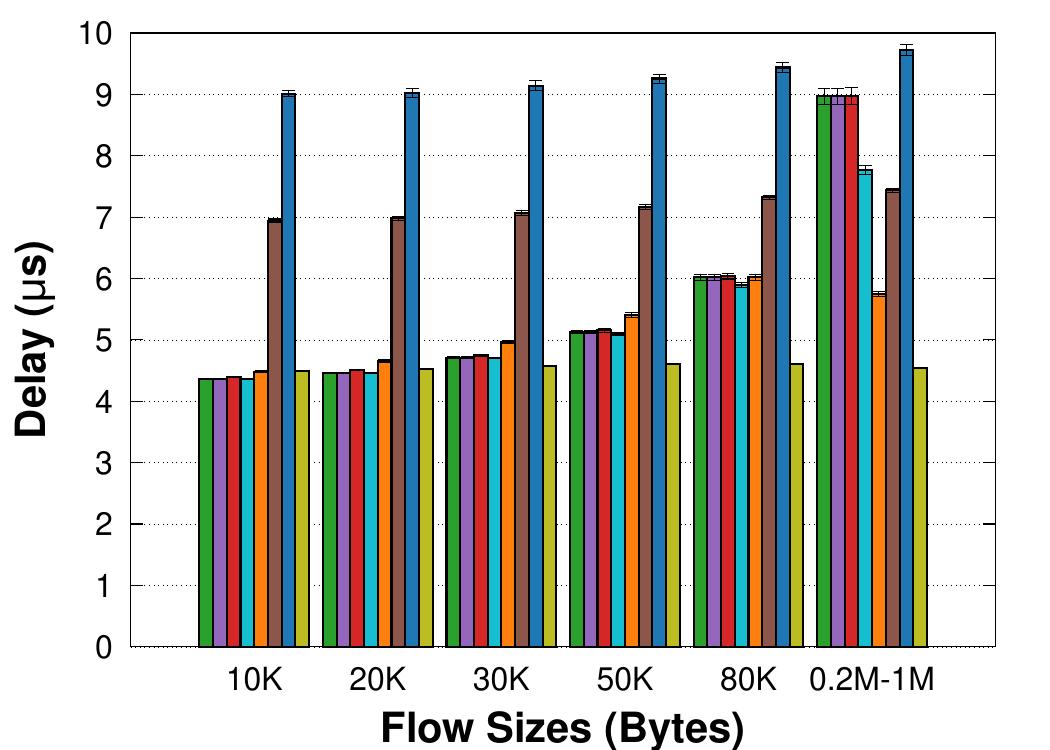}\label{vary-load-dc-wfq-delay-90}}
            \caption{Per packet delay for Datacenter topology with STFQ scheduling}
            \label{vary-load-dc-wfq-delay}
        \end{figure*}

\purpleblock{
    \subsection{PER-PACKET DELAY} \label{sect:perf-eval-delays}
        The per-packet delay of a packet from its source to destination is measured and shown in Fig.~\ref{vary-load-dc-pfabric-delay} and in Fig.~\ref{vary-load-dc-wfq-delay} for the Websearch workload. The simulation setup follows the same setup as used to measure the flow completion times in the previous section. In Fig.~\ref{vary-load-dc-pfabric-delay} with pFabric scheduling, we observe that the delay is consistent across different load levels for priority scheduling mechanisms. \acrshort{p3po} and \acrshort{ppqa} closely follow the ideal \acrshort{pifo} mechanism. Per-packet delay for \acrshort{sppifo} is also close to priority queuing mechanisms. This is primarily because, \acrshort{sppifo} uses multiple \acrshort{fifo} queues to separate priorities and thus can approximate \acrshort{pifo} behavior better than other approximate mechanisms with reduced packet drops. Approximate mechanisms like \acrshort{aifo} and \acrshort{rifo} have higher per-packet delay. This is due to the underlying queuing architecture used by these mechanisms, which is a single First-In First-Out, where once the packet is enqueued, its ordering cannot be changed. The use of a single \acrshort{fifo} causes head-of-the-line blocking. 
        
        For higher loads, \acrshort{rifo} performs slightly better than \acrshort{aifo} since it admits fewer packets compared to \acrshort{aifo}, leading to lower packet delay and thus lower head-of-the-line blocking. We observe, with 70~\% load, the delay is lower by 48.1~\% for small flows to 53.9~\% for larger flows. As the load increases, the improvement is from 67.9~\% for small flows to 70.7~\% for large flows. Fig.~\ref{vary-load-dc-wfq-delay} shows the per-packet delay when simulating with \acrshort{stfq} scheduling algorithm. The trends are similar to what we observe in pFabric. However, as the load increases, the per-packet delay for accurate schedulers increase for larger flows. This is because, multiple small flows arrive and finish before a large flow is finished. Thus, with packets arriving from every new small flow, the packets from larger flows are starved, leading to increased per-packet delay as they are increasingly scheduled later. 
        
        We also observe a surprising result here with the \acrshort{fifo} mechanism, since it behaves consistently similarly to the \acrshort{pifo} mechanism for smaller flows and has a lower per-packet delay for larger flows. Since \acrshort{fifo} relies on First-in First-out principle, and has no ordering based on packet priorities, it enqueues and dequeues packets according to their arrivals. Thus, this leads to some packets having higher delays and some packets having lower delays. Thus, we reason that these opposing delays nullify each leading to lower overall per-packet delays. 
}

\purpleblock{
    \subsection{THROUGHPUT} \label{sect:perf-eval-throughput}
        Fig.~\ref{vary-load-dc-pfabric-throughput} and Fig.~\ref{vary-load-dc-wfq-throughput} show the throughput between different flows for different load levels on the network. As seen in Fig.~\ref{vary-load-dc-pfabric-throughput} for pFabric scheduling, throughput directly correlates with \acrshort{fcts} as discussed in previous sections. The throughput of priority queuing mechanisms is lower compared to approximate queuing due to starvation of packets owing to pFabric's scheduling policy. However, with \acrshort{stfq} as seen in Fig.~\ref{vary-load-dc-wfq-throughput}, as there is no starvation, the throughput of \acrshort{p3po} and \acrshort{ppqa} is greater than the throughput of approximate schedulers. 
        As the load on the network increases, \acrshort{p3po} and \acrshort{ppqa} demonstrate an increasing trend in higher throughput compared to approximate schedulers. At 90\% load, \acrshort{p3po} and \acrshort{ppqa} perform 3\% to 33\% better for flows with size 10~KB and 16\% to 27\% for larger flows with size 200~KB-1~MB. 
        
        \begin{figure*}[htbp]
            \centering
            \includegraphics[width=0.9\linewidth]{journal-imgs/schedulers-legend-1rows.pdf}\\[-4ex]
            \subfloat[30\% load]{\includegraphics[width=0.32\linewidth]{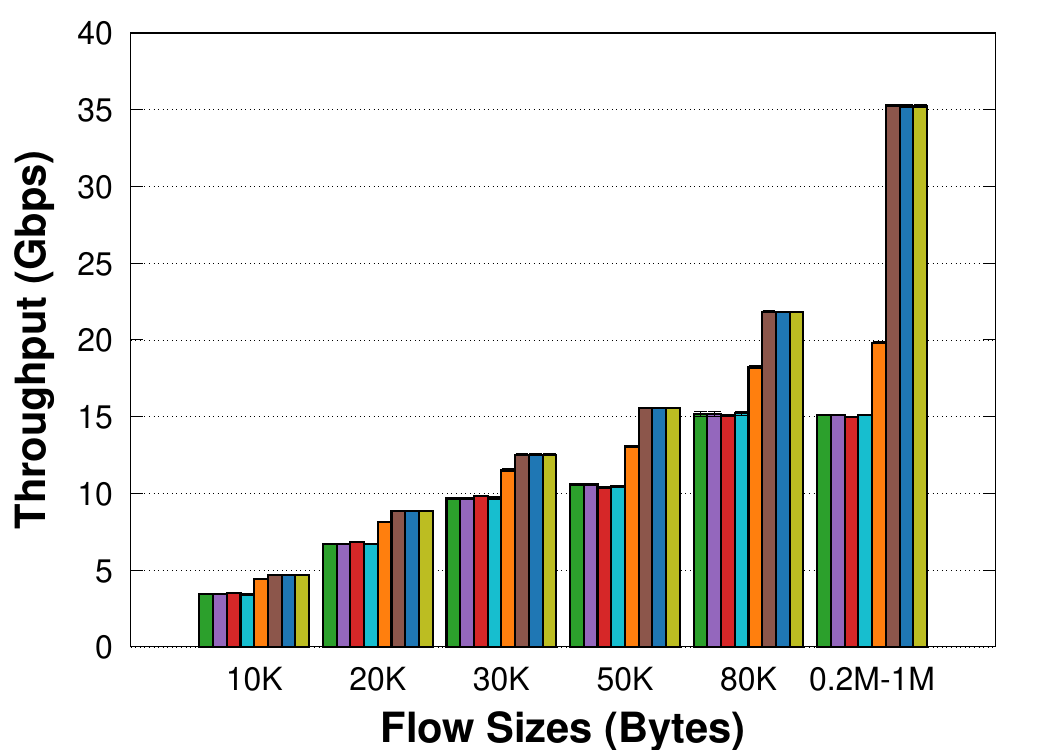}\label{vary-load-dc-pfabric-throughput-30}}
            \subfloat[70\% load]{\includegraphics[width=0.32\linewidth]{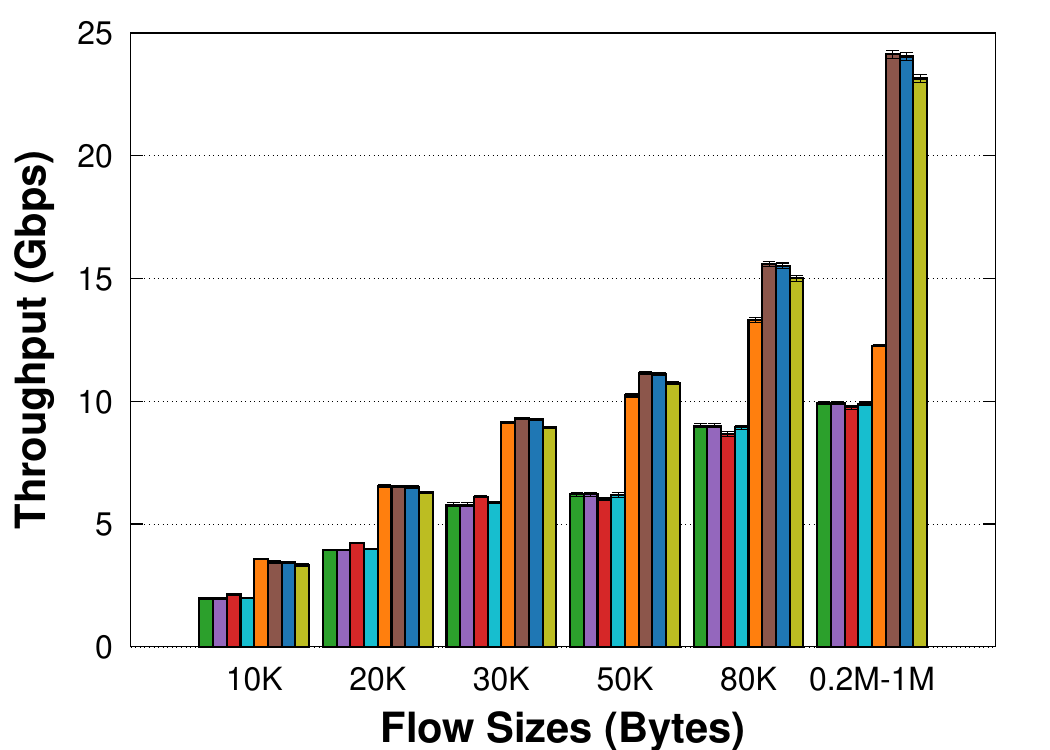}\label{vary-load-dc-pfabric-throughput-70}}
            \subfloat[90\% load]{\includegraphics[width=0.32\linewidth]{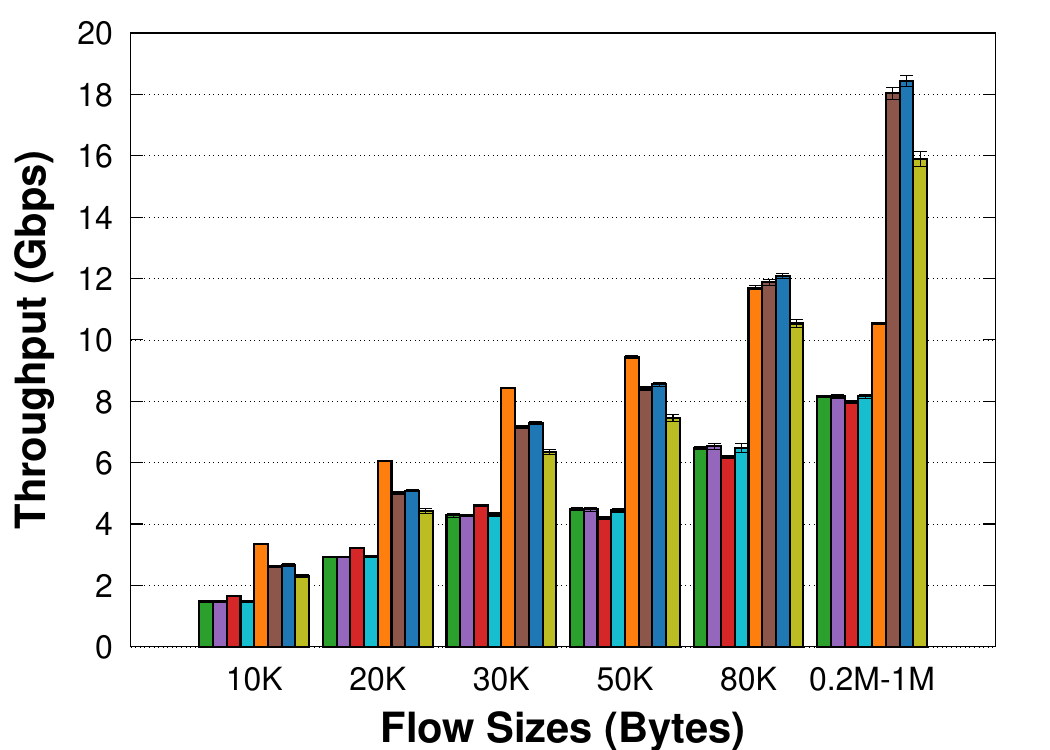}\label{vary-load-dc-pfabric-throughput-90}}
            \caption{Throughput for Datacenter topology with pFabric scheduling}
            \label{vary-load-dc-pfabric-throughput}
        \end{figure*}

        \begin{figure*}[htbp]
            \centering
            \includegraphics[width=0.9\linewidth]{journal-imgs/schedulers-legend-1rows.pdf}\\[-4ex]
            \subfloat[30\% load]{\includegraphics[width=0.32\linewidth]{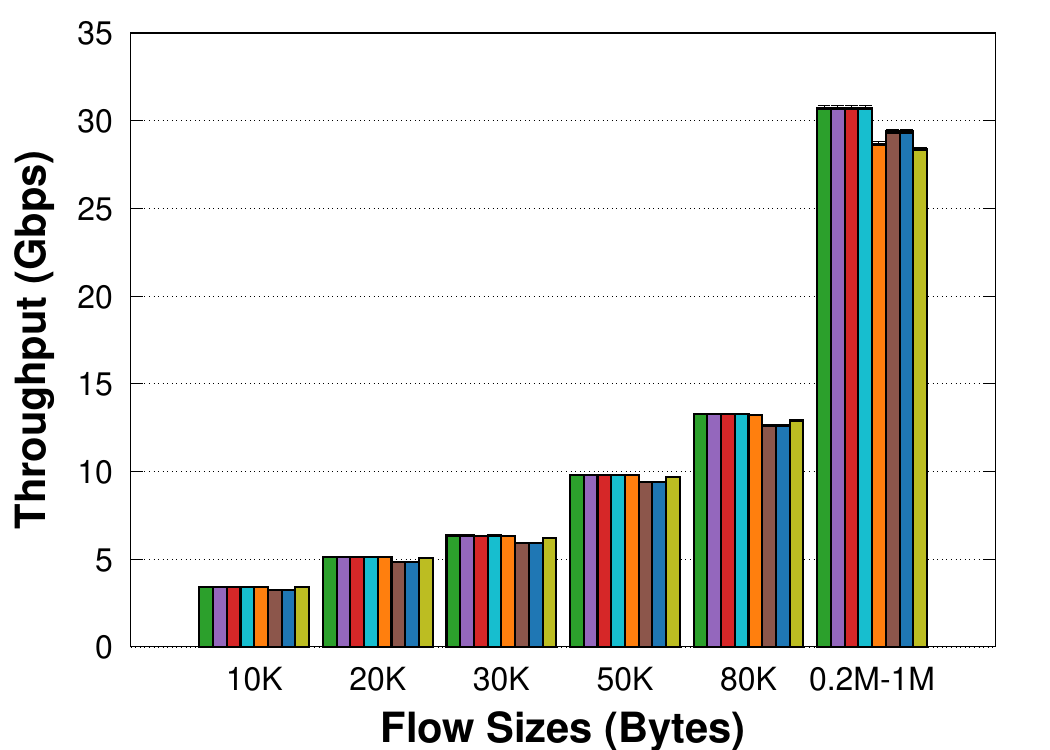}\label{vary-load-dc-wfq-throughput-30}}
            \subfloat[70\% load]{\includegraphics[width=0.32\linewidth]{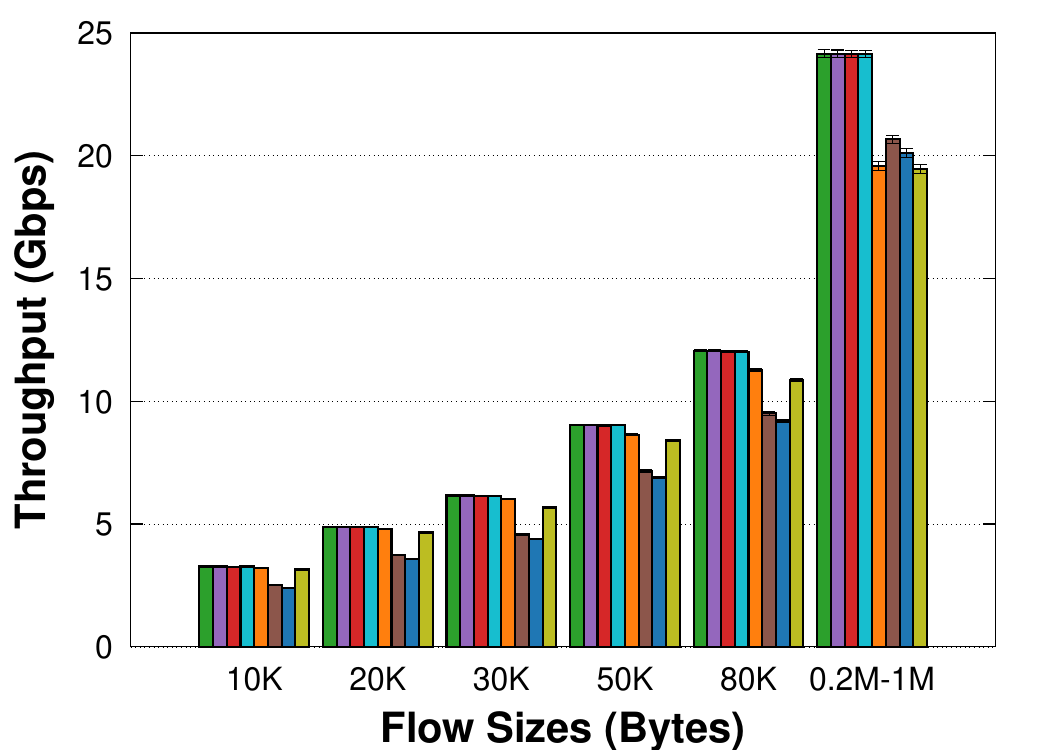}\label{vary-load-dc-wfq-throughput-70}}
            \subfloat[90\% load]{\includegraphics[width=0.32\linewidth]{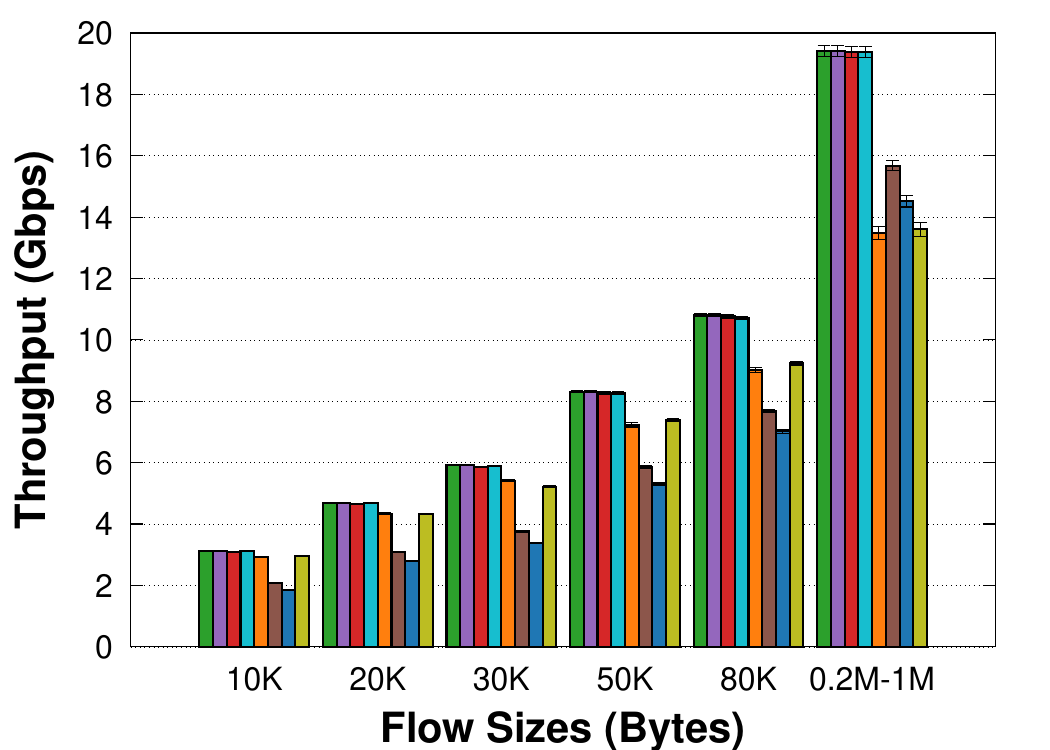}\label{vary-load-dc-wfq-throughput-90}}
            \caption{Throughput for Datacenter topology with STFQ scheduling}
            \label{vary-load-dc-wfq-throughput}
        \end{figure*}  
}

\blueblock{
    \subsection{FLOW METRICS FOR STAR TOPOLOGY} \label{sect:perf-eval-star}
        We also simulate the schedulers for a star topology with 64 hosts. The Fig~\ref{vary-load-STAR-wb} shows the flow metrics for a traffic load of 90\%. We observe similar trend in metrics as the datacenter topology. FCTs for pFabric remain higher than approximate schedulers owing to its starvation problem, but FCTs for STFQ are lower. The per-packet delays are lower for flows smaller than 80~KB. For larger flows, we observe that per-packet delays increase. This is because, smaller sized flows are generated frequently as compared to large flows, thus they need to be fairly allocated. This causes starvation for packets of larger flows, as new smaller flows are given priority.

        \begin{figure*}[htbp]
            \centering
            \includegraphics[width=0.9\linewidth]{journal-imgs/schedulers-legend-1rows.pdf}\\[-4ex]
            \subfloat[FCTs with pFabric]{\includegraphics[width=0.24\linewidth]{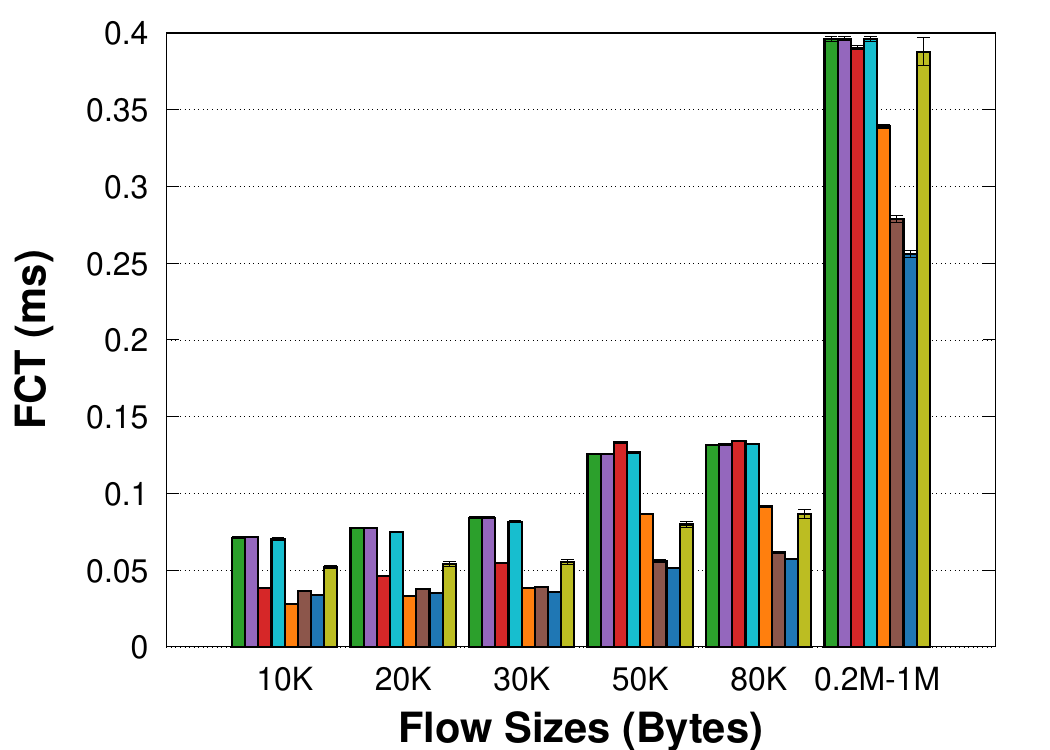}\label{vary-load-STAR-pfabric-wb-90-fct}}
            \subfloat[FCTs with STFQ]{\includegraphics[width=0.24\linewidth]{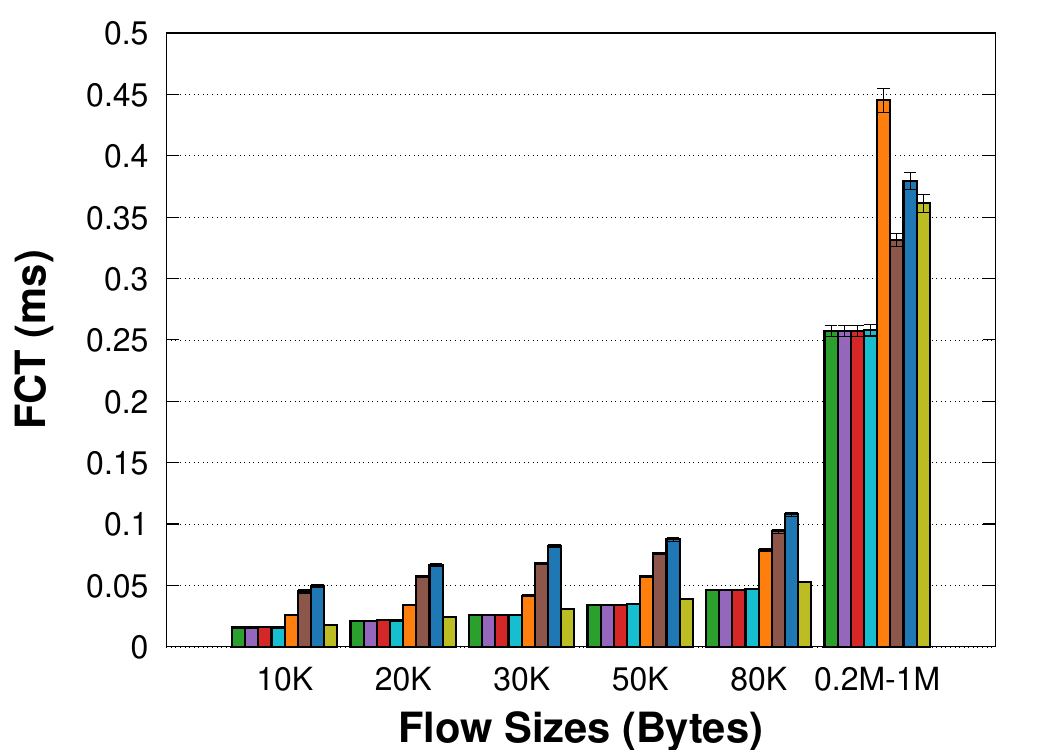}\label{vary-load-STAR-wfq-wb-90-fct}}
            \subfloat[Per-packet delay with pFabric]{\includegraphics[width=0.24\linewidth]{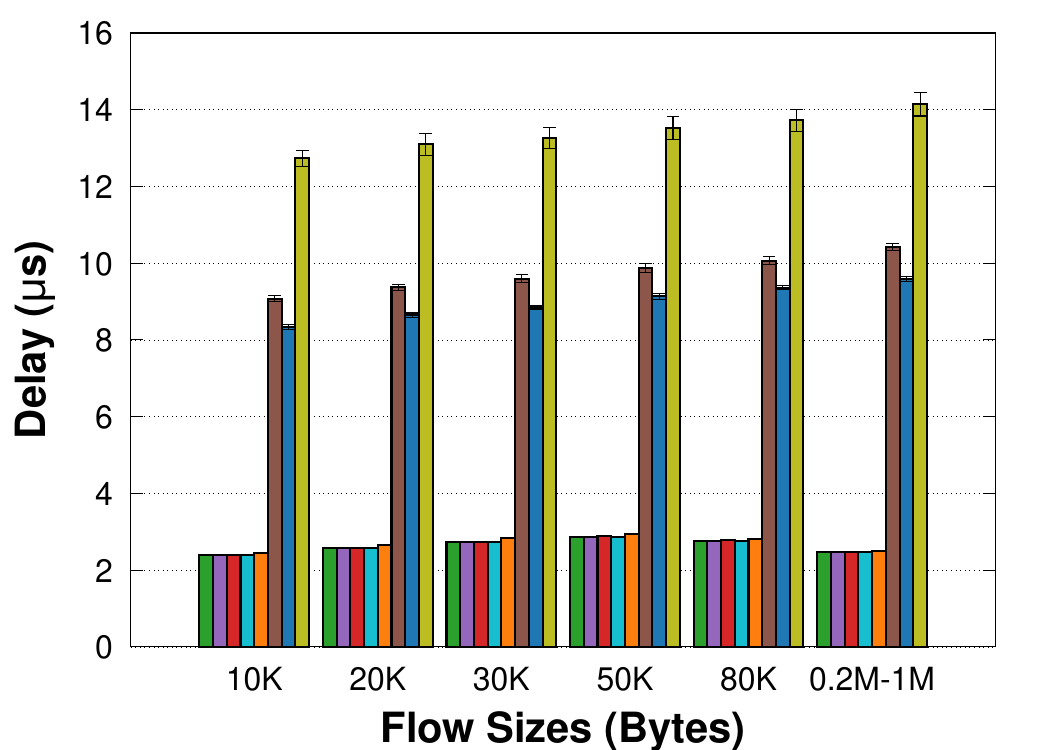}\label{vary-load-STAR-pfabric-wb-90-delay}}
            \subfloat[Per-packet delay with STFQ]{\includegraphics[width=0.24\linewidth]{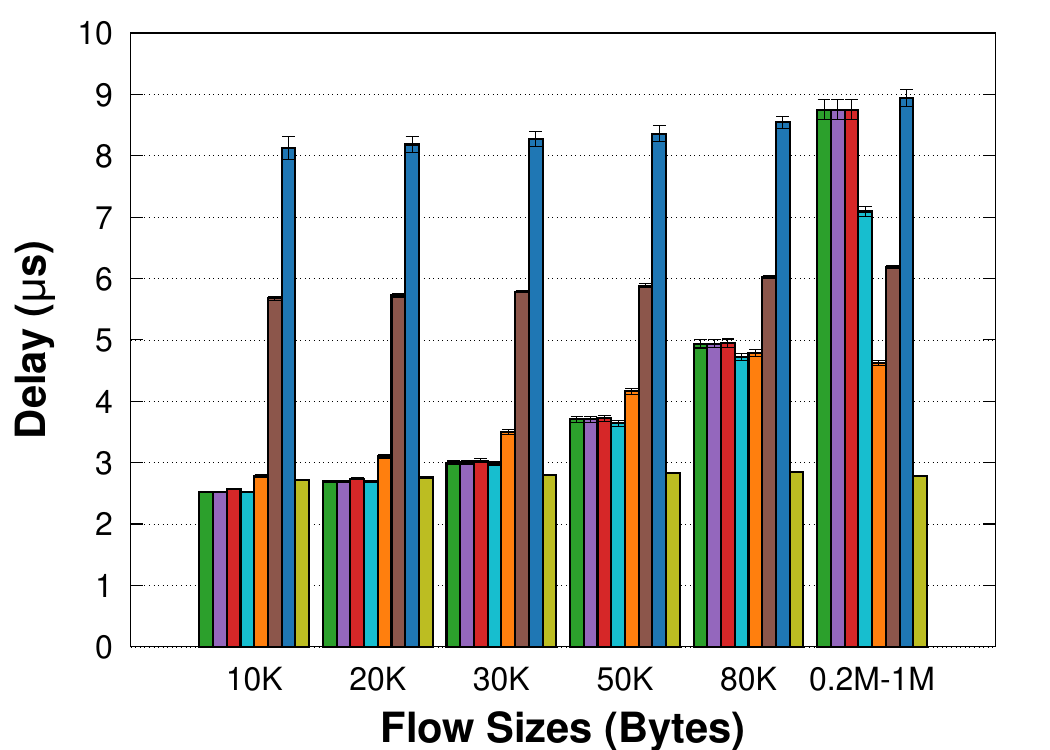}\label{vary-load-STAR-wfq-wb-90-delay}}\\[1ex]
            \caption{Evaluation on Star topology with 90\% load}
            \label{vary-load-STAR-wb}
        \end{figure*}
}

\blueblock{
    \subsection{DATAMINING WORKLOAD}  \label{sect:perf-eval-dm}
        Fig.~\ref{fig:vary-load-DC-wfq-dm} shows the \acrshort{fcts}, Per-packet delay and Throughput on the data center topology for \acrshort{stfq} scheduling with Datamining workload for 90~\% load. For \acrshort{stfq} as shown in Fig.~\ref{fig:vary-load-DC-wfq-dm-90-fct}, we observe lower \acrshort{fcts} for \acrshort{p3po} and \acrshort{ppqa}. As the load is higher, approximate schedulers have higher number of packet inversions. This leads to their higher FCTs. 
        As the Datamining workload is highly skewed, there are a large number of flows with smaller sizes, however most bytes are concentrated in larger flows which are less in number. Thus, we observe 85.1~\% lower \acrshort{fcts} for flows with sizes smaller than 3~KB and 58.5~\% lower \acrshort{fcts} for 10~KB flows.
        Similarly, as shown in Fig.~\ref{fig:vary-load-DC-wfq-dm-90-throughput}, we observe higher throughput, which is a direct consequence of having lower FCTs. The approximate techniques based on admission control strategy (AIFO, RIFO) demonstrate lower throughput than FIFO since they drop packets if they do not cross a certain threshold. 
        \begin{figure}[htbp]
            \centering
            \includegraphics[width=\linewidth]{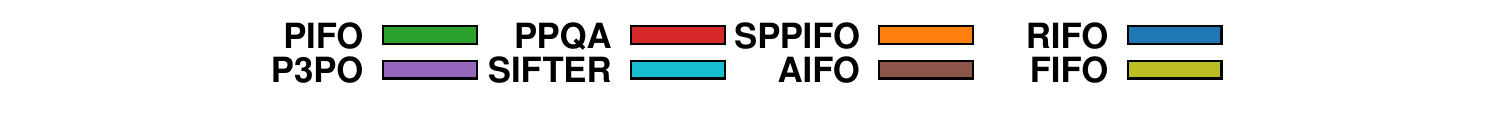}\\[-4ex]
            
            \subfloat[FCTs]{\includegraphics[width=0.48\linewidth]{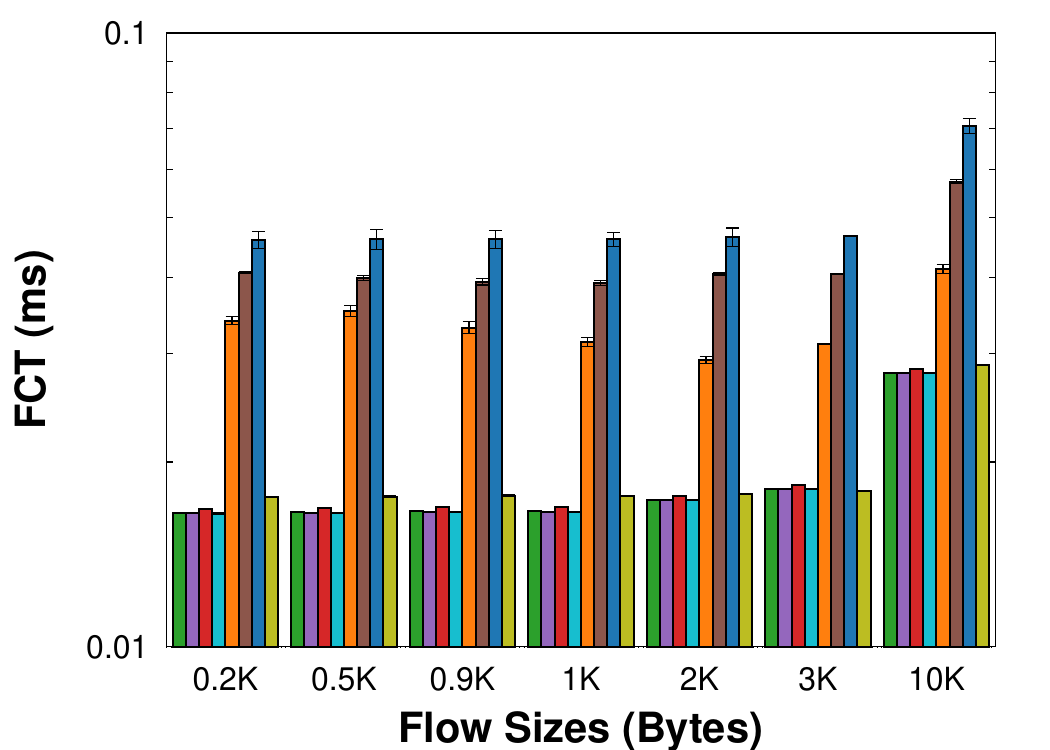}\label{fig:vary-load-DC-wfq-dm-90-fct}}
            \subfloat[Throughput]{\includegraphics[width=0.48\linewidth]{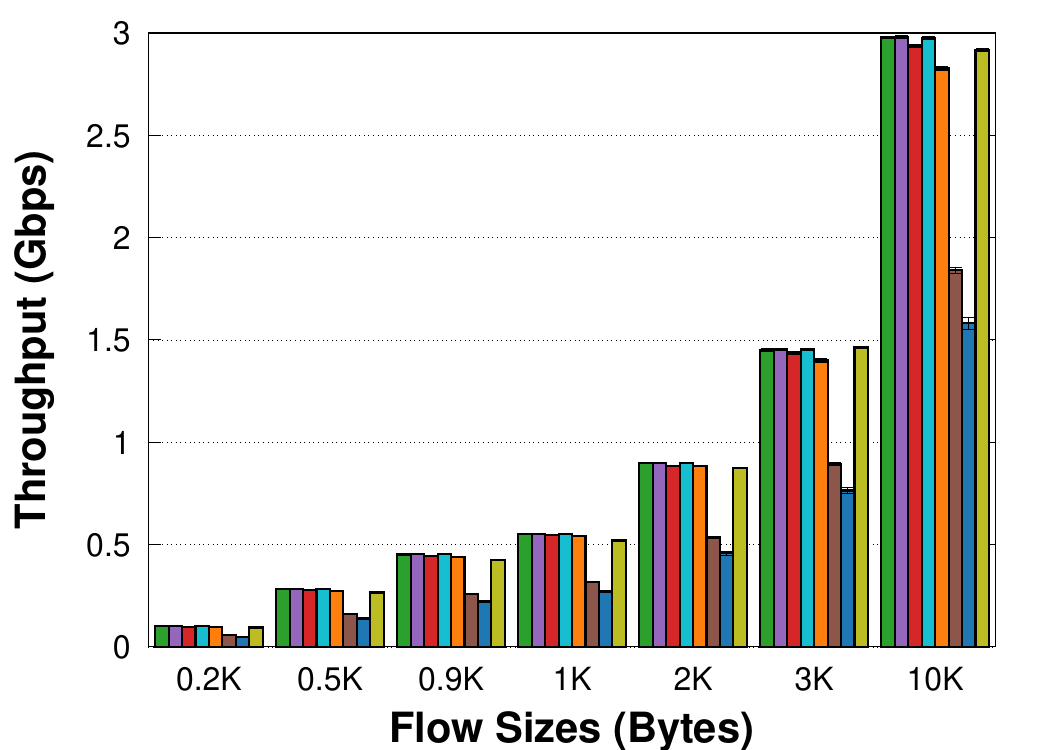}\label{fig:vary-load-DC-wfq-dm-90-throughput}}
            
            \caption{Evaluation on data center topology with STFQ scheduling and Datamining workload for 90\% load}
            \label{fig:vary-load-DC-wfq-dm}
        \end{figure}
}

\section{CONCLUSIONS} \label{sect:conclusions}
    This paper presented two parallel architectures \acrshort{p3po} and \acrshort{ppqa}, scalable, inversion-free programmable packet scheduling architectures based on \acrshort{pifo}-abstraction queues that addresses the scalability and sorting limitations of traditional programmable schedulers. \acrshort{p3po} and \acrshort{ppqa} partition a global priority queue into multiple smaller, independently sorted queues. These architectures maintain the strict priority ordering required by custom scheduling policies. Using a simulation model developed using the NetBench platform, we  showed that the performance of \acrshort{p3po} and \acrshort{ppqa} was comparable to the ideal \acrshort{pifo}, in terms of fair bandwidth allocation, comparable flow completion times, and per-packet delays under different-load levels in data-center workloads. \acrshort{p3po} and \acrshort{ppqa} was also compared to Sifter, an inversion-free packet scheduler, and \acrshort{sppifo}, \acrshort{aifo} and \acrshort{rifo}, approximate schedulers. \acrshort{p3po} and \acrshort{ppqa} offers simplicity in implementation by avoiding Sifter's complex control loop of moving packets between the \acrshort{fifo}s and priority queue. It also avoids the packet inversion errors introduced by approximate schedulers. As a future work, \acrshort{p3po} and \acrshort{ppqa} can be extended to support hierarchical priority queuing, \acrshort{aqm} techniques that allow more complex scheduling hierarchies and multi-level QoS enforcement.

\section*{Acknowledgments}
This work was supported by a research grant from Ciena Corporation. 

\bibliographystyle{IEEEtran}
\bibliography{ref}





\end{document}